\documentclass[journal,compsoc]{IEEEtran}

\newif\ifincludeappendix
\includeappendixtrue 

\usepackage{tikz}
\usetikzlibrary{arrows.meta}
\usetikzlibrary{fit}

\usepackage{hyperref}
\usepackage{url}
\usepackage{booktabs}
\usepackage{fdsymbol} 
\usepackage{tipa} 
\usepackage{amsthm}
\usepackage{cleveref}
\usepackage{subcaption}
\usepackage{enumerate}

\RequirePackage[noline,noend,linesnumbered]{algorithm2e}
\crefname{algocf}{Algorithm}{Algorithms}
\newcommand\auxfun[1]{\expandafter\newcommand\csname #1\endcsname{%
 \mathop{\hbox{$\mathsf{#1}$}}\nolimits}}
\newcommand\af[1]{\mathop{\hbox{$\mathsf{#1}$}}\nolimits}
\newcommand{\V}{\mb{V}}

\newtheorem{theorem}{Theorem}
\newtheorem{definition}[theorem]{Definition}
\newtheorem{lemma}[theorem]{Lemma}

\newcommand{\msf}[1]{\ensuremath{\mathsf{#1}}}
\newcommand{\mb}[1]{\ensuremath{\mathbb{#1}}}

\newcommand{\N}{\mb{N}}

\begin{document}
\title{The Art of the Fugue: Minimizing Interleaving in Collaborative Text Editing}
\author{Matthew Weidner and Martin Kleppmann\IEEEcompsocitemizethanks{%
    \IEEEcompsocthanksitem M.\ Weidner is with Carnegie Mellon University. %
    \IEEEcompsocthanksitem M.\ Kleppmann is with the University of Cambridge.}}

\IEEEtitleabstractindextext{%
\begin{abstract}
Most existing algorithms for replicated lists, which are widely used in collaborative text editors, suffer from a problem: when two users concurrently insert text at the same position in the document, the merged outcome may interleave the inserted text passages, resulting in corrupted and potentially unreadable text.
The problem has gone unnoticed for decades, and it affects both CRDTs and Operational Transformation.
This paper defines maximal non-interleaving, our new correctness property for replicated lists.
We introduce two related CRDT algorithms, Fugue and FugueMax, and prove that FugueMax satisfies maximal non-interleaving.
We also implement our algorithms and demonstrate that Fugue offers performance comparable to state-of-the-art CRDT libraries for text editing.
\end{abstract}

}
\maketitle

\IEEEraisesectionheading{\section{Introduction}}
\label{sec:intro}

\IEEEPARstart{C}{ollaborative} text editors such as Google Docs allow several users to concurrently modify a document, while ensuring that all users' copies of the document converge to the same state.
Even though algorithms for collaborative text editing have been studied for over three decades~\cite{Ellis:1989,Nichols:1995,Sun:1998:tochi}, a formal specification of the required behavior only appeared as recently as 2016~\cite{attiya}.
We argue that this specification is incomplete.
There is an additional correctness property that is important in practice, but which has been overlooked by almost all prior research on this topic: \emph{non-interleaving}.
Informally stated, this property requires that when sections of text are composed independently from each other (perhaps while the users are offline), and the edits are subsequently merged, those sections are placed one after another, and not intermingled in the final document. 

For example, suppose two users are editing a text document containing a shopping list, as shown in \Cref{fig:forward-interleaving}. Initially, the document contains the word ``milk'' and a newline character. User A inserts a line break and ``eggs'', while concurrently (due to a weak network connection), user B inserts a line break and ``bread''. A collaborative text editor that is correct with respect to the existing formal specification may choose to merge their edits as shown: ``milk'', a blank line, and then the interleaved word ``ebgrgesad''.

\begin{figure}\centering
  \begin{tikzpicture}[font=\footnotesize]
	\tikzstyle{state}=[matrix,column sep={15pt,between origins}]
	\tikzstyle{val}=[draw,anchor=base,minimum width=15pt,text height=6pt,text depth=2pt,inner xsep=0]
	\tikzstyle{oid}=[anchor=base,font=\tiny]
	\tikzstyle{ins1}=[fill=red!10]
	\tikzstyle{ins2}=[fill=blue!10]
	\tikzstyle{network}=[thick,-{Stealth[length=3mm]}]
	\node (hello) at (1.5,3.3) [state] {
		\node [val] {\verb+m+}; &
		\node [val] {\verb+i+}; &
		\node [val] {\verb+l+}; &
		\node [val] {\verb+k+}; &
		\node [val] {\verb+\n+}; \\
	};
	\node (change1) at (0,2.4) [state] {
		\node [val] {\verb+m+}; &
		\node [val] {\verb+i+}; &
		\node [val] {\verb+l+}; &
		\node [val] {\verb+k+}; &
        \node [val,ins1] {\verb+\n+}; &
        \node [val,ins1] {\verb+e+}; &
        \node [val,ins1] {\verb+g+}; &
        \node [val,ins1] {\verb+g+}; &
        \node [val,ins1] {\verb+s+}; &
		\node [val] {\verb+\n+}; \\
	};
	\node (change2) at (3,1.5) [state] {
		\node [val] {\verb+m+}; &
		\node [val] {\verb+i+}; &
		\node [val] {\verb+l+}; &
		\node [val] {\verb+k+}; &
        \node [val,ins2] {\verb+\n+}; &
        \node [val,ins2] {\verb+b+}; &
        \node [val,ins2] {\verb+r+}; &
        \node [val,ins2] {\verb+e+}; &
        \node [val,ins2] {\verb+a+}; &
        \node [val,ins2] {\verb+d+}; &
		\node [val] {\verb+\n+}; \\
	};
	\node (interleaved) at (1.5,0.3) [state] {
		\node [val] {\verb+m+}; &
		\node [val] {\verb+i+}; &
		\node [val] {\verb+l+}; &
		\node [val] {\verb+k+}; &
        \node [val,ins1] {\verb+\n+}; &
        \node [val,ins2] {\verb+\n+}; &
        \node [val,ins1] {\verb+e+}; &
        \node [val,ins2] {\verb+b+}; &
        \node [val,ins1] {\verb+g+}; &
        \node [val,ins2] {\verb+r+}; &
        \node [val,ins1] {\verb+g+}; &
        \node [val,ins2] {\verb+e+}; &
        \node [val,ins1] {\verb+s+}; &
        \node [val,ins2] {\verb+a+}; &
        \node [val,ins2] {\verb+d+}; &
		\node [val] {\verb+\n+}; \\
	};
    \node at (change1.north west) [above right] {User A:};
    \node at (change2.north east) [above left] {User B:};
    \node at (interleaved.north west) [above right] {merged:};
	\draw [network] (hello.west)  to [out=180,in=70] (change1.130);
	\draw [network] (hello.east)  to [out=0,in=90]   (change2.50);
	\draw [network] (change1.230) to [out=270,in=130] (interleaved.164);
	\draw [network] (change2.310) to [out=240,in=40] (interleaved.16);
  \end{tikzpicture}
  \caption{Possible interleaving in a collaborative text editor.}
  \label{fig:forward-interleaving}
\end{figure}

Several existing text collaboration algorithms interleave text in this way, and almost all algorithms that we surveyed interleave text in some cases. 
Affected algorithms include both main approaches to collaborative text editing: Conflict-free Replicated Data Types (CRDTs) \cite{Shapiro:2011, crdt_summary_2018}, and Operational Transformation (OT)~\cite{Ellis:1989,Sun:1998:tochi}.

This paper describes the interleaving problem in detail and introduces novel algorithms that minimize interleaving. 
One surprising result of this paper is that it is impossible to avoid interleaving in every situation. Users can perform multiple interacting concurrent updates in such a way that \emph{any} algorithm must interleave some characters in the merged text, at least according to a naive definition that forbids interleaving both forward and backward insertions (defined in \Cref{sec:backward-interleaving}). Thus we instead define a new correctness property, \emph{maximal non-interleaving}, 
which forbids interleaving to a maximum possible extent.

We then introduce \emph{Fugue} and \emph{FugueMax}, two novel CRDT algorithms for collaborative text editing. We prove that both algorithms avoid interleaving text except in the rare situations where some interleaving is inevitable. Specifically, FugueMax satisfies maximal non-interleaving, 
while Fugue is simpler but may interleave more characters than necessary in situations where interleaving is inevitable. 

Finally, we provide an optimized open source implementation of Fugue and show that it achieves performance comparable to the state-of-the-art Yjs library on a realistic text-editing trace. 


\section{Background}
\label{sec:background}

CRDT and OT algorithms for text allow multiple users, each with their own copy of the document, to concurrently edit the text.
Although these algorithms are usually presented in the context of text editing, they can easily be generalized beyond text: instead of a list of characters, the algorithm could manage a list of other objects, such as items on a to-do list, or rows in a spreadsheet.
We therefore also say that these algorithms implement a replicated list object~\cite{attiya}.

\subsection{System Model}

In collaborative text editors, each user session (e.g., a tab in a web browser) maintains a replica of the list of characters.
On user input, the user's local replica of the document is updated by inserting or deleting characters in this list, and each such insertion or deletion is called an \emph{operation}.
A user's operations are immediately applied to their local replica, without waiting for network communication with any other nodes, in order to provide responsive user interaction independently of network latency.
This model also allows disconnected operation: if a user edits the document while offline, their client buffers the operations they generate and sends them to collaborators when they next connect.

Assuming users eventually come online, every operation is eventually propagated to other replicas, and each replica integrates remote operations into its local state as they are received.
At a minimum, this process must ensure \emph{convergence}: any two replicas that have processed the same set of operations must be in the same state, even if they received the operations in a different order.
A more detailed specification of a replicated list is given by Attiya at al.~\cite{attiya}; we summarize it in the proof of \Cref{thm:correctness} in \Cref{sec:tree_fugue}.

Algorithms for collaborative text editing differ in the assumptions they make about the network between replicas.
For example, the OT algorithm Jupiter~\cite{Nichols:1995} assumes that all operations pass through a central server that sequences and transforms those operations.
On the other hand, CRDTs (including our Fugue algorithm in \Cref{sec:tree_fugue}) generally assume a causal broadcast protocol~\cite{Shapiro:2011}.
Causal broadcast can be implemented in a peer-to-peer network without assuming any central server or consensus protocol~\cite{Birman:1991,Cachin:2011}, making this model suitable for decentralized systems.
The causal broadcast protocol handles retransmission of dropped messages and ensures that when a replica comes online, it receives all the operations it missed while offline. 


\subsection{The interleaving problem}\label{sec:interleaving-problem}

Several replicated list CRDTs, including Treedoc~\cite{treedoc}, Logoot~\cite{logoot}, and LSEQ \cite{lseq}, assign to each list element a unique identifier from a dense, totally ordered set.
The sequence of list elements is then obtained by sorting the IDs in ascending order.
To insert a new list element between two adjacent elements with IDs $\mathit{id}_1$ and $\mathit{id}_2$ respectively, the algorithm generates a new unique ID $\mathit{id}_3$ such that $\mathit{id}_1 < \mathit{id}_3 < \mathit{id}_2$, where $<$ is the total order on identifiers.

Say another user concurrently inserts an element with ID $\mathit{id}_4$ between the same pair of elements $(\mathit{id}_1, \mathit{id}_2)$ such that $\mathit{id}_1 < \mathit{id}_4 < \mathit{id}_2$.
The minimum requirement of $\mathit{id}_3 \neq \mathit{id}_4$ is easy to achieve (e.g., by including in each ID the unique name of the replica that generated it), but whether $\mathit{id}_3 < \mathit{id}_4$ or $\mathit{id}_3 > \mathit{id}_4$ is an arbitrary choice.

When two users concurrently insert several new elements in the same ID interval, the result is the effect illustrated in the introduction's \Cref{fig:forward-interleaving}. For example, in the Treedoc list CRDT, each ID is a path through a binary tree together with a tiebreaker. When two users insert at the same location, they choose the same paths through the binary tree: `e' in ``eggs'' and `b` in ``bread'' use the same path, as do `g` in ``eggs'' and `r` in ``bread'', etc. The algorithm then visits each path in tree traversal order; with tiebreakers, the merged result thus alternates between the letters of each word: ``ebgrgesad''.

We argue that this behavior is obviously undesirable.
Nevertheless, none of the affected papers even mention the issue, and although it had been informally known to people in the field for some time, it was not documented in the research literature until 2018 \cite{opsets,Sun:2018}.
Attiya et al.'s specification~\cite{attiya} allows interleaving like in \Cref{fig:forward-interleaving}.

\subsection{Interleaving of forward and backward insertions}
\label{sec:backward-interleaving}
In some replicated list algorithms, whether interleaving can occur or not depends on the order in which the elements are inserted into the list.
In the common case, when a user writes text, they insert characters in forward direction: that is, ``bread'' is inserted as the character sequence ``b'', ``r'', ``e'', ``a'', ``d''.
However, not all writing is in forward direction: sometimes users hit backspace to fix a typo, or move their cursor to a different location in the document and continue typing there.

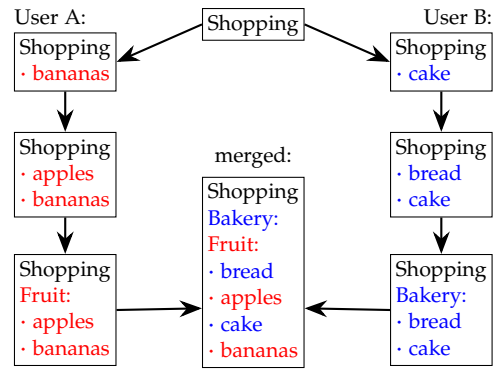
\begin{figure}\centering
  \begin{tikzpicture}[font=\footnotesize]
    \tikzstyle{box}=[rectangle,draw,matrix,text height=6pt,text depth=2pt,nodes=right,inner sep=1pt]
    \tikzstyle{network}=[thick,-{Stealth[length=3mm]}]
    \node (start) at (2.5,4) [box] {
        \node {Shopping}; \\
    };
    \node (left1) at (0,3.5) [box] {
        \node {Shopping}; \\
        \node [red] {$\cdot$ bananas}; \\
    };
    \node (right1) at (5,3.5) [box] {
        \node {Shopping}; \\
        \node [blue] {$\cdot$ cake}; \\
    };
    \node (left2) at (0,2) [box] {
        \node {Shopping}; \\
        \node [red] {$\cdot$ apples}; \\
        \node [red] {$\cdot$ bananas}; \\
    };
    \node (right2) at (5,2) [box] {
        \node {Shopping}; \\
        \node [blue] {$\cdot$ bread}; \\
        \node [blue] {$\cdot$ cake}; \\
    };
    \node (left3) at (0,0.2) [box] {
        \node {Shopping}; \\
        \node [red] {Fruit:}; \\
        \node [red] {$\cdot$ apples}; \\
        \node [red] {$\cdot$ bananas}; \\
    };
    \node (right3) at (5,0.2) [box] {
        \node {Shopping}; \\
        \node [blue] {Bakery:}; \\
        \node [blue] {$\cdot$ bread}; \\
        \node [blue] {$\cdot$ cake}; \\
    };
    \node (merged) at (2.5,0.7) [box] {
        \node {Shopping}; \\
        \node [blue] {Bakery:}; \\
        \node [red] {Fruit:}; \\
        \node [blue] {$\cdot$ bread}; \\
        \node [red] {$\cdot$ apples}; \\
        \node [blue] {$\cdot$ cake}; \\
        \node [red] {$\cdot$ bananas}; \\
    };
    \node at (left1.north west) [above right, inner xsep=0, inner ysep=3pt] {User A:};
    \node at (right1.north east) [above left, inner xsep=0, inner ysep=3pt] {User B:};
    \node at (merged.north) [above] {merged:};
    \draw [network] (start.west) -- (left1.east);
    \draw [network] (start.east) -- (right1.west);
    \draw [network] (left1.south) -- (left2.north);
    \draw [network] (right1.south) -- (right2.north);
    \draw [network] (left2.south) -- (left3.north);
    \draw [network] (right2.south) -- (right3.north);
   	\draw [network] (left3.east) -- (merged.215);
    \draw [network] (right3.west) -- (merged.325);
  \end{tikzpicture}
  \caption{Each user prepends items to their shopping list. When the edits are merged in an algorithm that allows interleaving of backward insertions, the result may place the items in an illogical order, such as listing bread in the fruit category.}
  \label{fig:backward-interleaving}
\end{figure}

Besides inserting characters in forward direction, it is also possible to insert in backward direction.
The extreme case of typing text in reverse order character by character (typing ``bread'' as ``d'', ``a'', ``e'', ``r'', ``b'') is unlikely to occur in practical text editing scenarios.
However, a plausible scenario of backward insertion is illustrated in \Cref{fig:backward-interleaving}.
In this example, two users add items to a shared shopping list while offline.
Each user prepends new items to the beginning of the list, finally adding a category header (``Fruit'' or ``Bakery'') for the items they added.
When the users merge their changes, an algorithm that allows interleaving of backward insertions may place the items in a surprising order.
This behavior is less bad than the fine-grained character-by-character interleaving of \Cref{fig:forward-interleaving}, but it is nevertheless not ideal.
It would be preferable to keep all of each user's insertions as one contiguous string, regardless of the order in which the elements were inserted.

When OT/CRDT algorithms for replicated lists are used for data other than text, backward insertion is more likely to occur.
For example, in a spreadsheet or to-do list, new rows/items might regularly be inserted at the top, one at a time.
If we can avoid both forward and backward interleaving, we also improve the behavior of these applications.

\section{Related Work}
\label{sec:related}

Collaborative text editing originated with the work of Ellis and Gibbs~\cite{Ellis:1989}, who also introduced Operational Transformation (OT) as a technique for resolving concurrent edits.
This approach was formalized by Ressel et al.~\cite{Ressel:1996}, and further developed by Sun et al.~\cite{Sun:1998:cscw} and many other papers.
The OT algorithm Jupiter~\cite{Nichols:1995} later became the basis for real-time collaboration in Google Docs~\cite{DayRichter:2010}.
Following bugs in several OT algorithms, which failed to converge in some situations~\cite{Imine:2003,Oster:2005}, Conflict-free Replicated Data Types (CRDTs) were developed as an alternative approach~\cite{Shapiro:2011}.
The first CRDT for text editing was WOOT~\cite{woot}, which was followed by Treedoc~\cite{treedoc}, Logoot~\cite{logoot}, RGA~\cite{rga}, and others.

\subsection{Algorithms that exhibit interleaving}
\label{sec:survey}
The interleaving problem was first noticed in CRDTs such as Logoot and LSEQ because they are particularly prone to the problem; experiments with implementations of these algorithms are easily able to trigger interleaving in practice~\cite{InterleavingTests}.
However, when we started looking at the issue more closely, we found that interleaving is surprisingly prevalent among both OT and CRDT algorithms for collaborative text editing.
Our findings are summarized in \Cref{tab:survey}, and examples of each instance of interleaving are detailed in \ifincludeappendix Appendix~\ref{sec:interleaving-examples}. \else the appendix. \fi

\begin{table}
    \centering
    \caption{Various algorithms' susceptibility to interleaving anomalies. Key: $\medblackcircle$ = interleaving can occur; $\medcircle$ = we have not been able to find examples of interleaving; $\medcircle\checkmark$ = proven not to interleave; $\lightning$ = algorithm may incorrectly reorder characters. Examples of anomalies appear in \ifincludeappendix Appendix~\ref{sec:interleaving-examples}.\else the appendix.\fi}\label{tab:survey}
    \renewcommand{\arraystretch}{1.2}
    \begin{tabular}{llllll} \toprule
       Family & Algorithm &
       \multicolumn{1}{p{1.3cm}}{forward \mbox{interleaving} (one replica)} &
       \multicolumn{1}{p{1.4cm}}{backward \mbox{interleaving} (one replica)} &
       \multicolumn{1}{p{1.4cm}}{backward \mbox{interleaving} (multi-replica)}
       \\\midrule
       OT & adOPTed \cite{Ressel:1996} & $\medblackcircle$ & $\medcircle$ & $\medblackcircle$ \\
       & Jupiter \cite{Nichols:1995} & $\medblackcircle$ & $\medcircle$ & $\medcircle$ \\
       & GOT \cite{Sun:1998:tochi} & $\medblackcircle$ & $\lightning$ & $\lightning$ \\
       & SOCT2 \cite{Suleiman:1997} & $\medblackcircle$ & $\medblackcircle$ & $\medblackcircle$ \\
       & TTF \cite{Oster:2006} & $\medblackcircle$ & $\medcircle$ & $\medblackcircle$ \\ \addlinespace[5pt]
       CRDT & WOOT \cite{woot} & $\medblackcircle$ & $\medcircle$ & $\medcircle$ \\
       & Logoot \cite{logoot} & $\medblackcircle$ & $\medblackcircle$ & $\medblackcircle$ \\
       & LSEQ \cite{lseq} & $\medblackcircle$ & $\medblackcircle$ & $\medblackcircle$ \\
       & Treedoc \cite{treedoc} & $\medblackcircle$ & $\medblackcircle$ & $\medblackcircle$ \\
       & RGA \cite{rga} & $\medcircle\checkmark$ & $\medblackcircle$ & $\medblackcircle$ \\
       & Yjs \cite{yjs} & $\medcircle\checkmark$ & $\medcircle$ & $\medblackcircle$ \\
       & Sync9 \cite{Sync9} & $\medcircle$ & $\medcircle$ & $\medcircle$ \\
       & YjsMod \cite{gentle_code} & $\medcircle$ & $\medcircle$ & $\medcircle$ \\
       & Fugue & $\medcircle\checkmark$ & $\medcircle\checkmark$ & $\medcircle\checkmark$ \\
       & FugueMax & $\medcircle\checkmark$ & $\medcircle\checkmark$ & $\medcircle\checkmark$
       \\\bottomrule
    \end{tabular}
\end{table}

Occurrence of interleaving is often nondeterministic, and the probability of exhibiting interleaving varies depending on the algorithm: for example, in some algorithms it depends on the exact order in which concurrently sent network messages are received, and in some it depends on random numbers generated as part of the algorithm.

In some algorithms, interleaving occurs only if multiple replicas participate in one of the concurrent editing sessions; this is indicated in the column labeled ``multi-replica'' in \Cref{tab:survey}.
This can happen, for example, if a user starts some work on one device and then continues on another device (producing an editing session that spans two devices), while independently another user works offline on the same document on a third device.
It can also occur in systems with ephemeral replica IDs, such as a web application that generates a fresh ID for every browser tab refresh.

In the cases marked $\medcircle$ in \Cref{tab:survey} we conjecture non-interleaving, but we have not proved it.
Only in the cases marked $\medcircle\checkmark$ has non-interleaving been proven.
RGA forward non-interleaving was proved by Kleppmann et al.~\cite{opsets}, Yjs forward non-interleaving is proved in an in-progress paper, and our own algorithms Fugue and FugueMax are verified in \Cref{sec:interleaving}. (For the table's claims of backward non-interleaving, we exclude situations where some interleaving is inevitable for any algorithm; see \Cref{sec:maximal}.)

The only existing algorithms for which we have not found interleaving examples are Sync9~\cite{Sync9}, and a modification of Yjs proposed by Seph Gentle (``YjsMod'')~\cite{gentle_code}.
At the time of writing, there are no research papers describing these algorithms, and no proofs of correctness; the only published material is source code and informal documentation~\cite{Sync9,gentle_code}.
Fugue was developed independently from both of these algorithms, while FugueMax uses a technique from YjsMod (see \Cref{sec:fugue_max_alg}).
We conjecture that Sync9 is semantically equivalent to Fugue, while YjsMod is semantically equivalent to FugueMax. If that is indeed the case, then Sync9 and YjsMod satisfy forward and backward non-interleaving.


\subsection{Previous attempt to ensure non-interleaving}
\label{sec:papoc-paper}

Kleppmann et al.~\cite{interleaving} previously identified the interleaving problem.
That work has two serious flaws:
\begin{enumerate}
    \item The definition of non-interleaving in that paper cannot be satisfied by any algorithm.
    \item The CRDT algorithm proposed in that paper, which aims to be non-interleaving, is incorrect~-- it does not converge.
    An example of non-convergence, which was found by Chandrassery~\cite{adithya}, is given in \ifincludeappendix Appendix~\ref{sec:adithya-bug}.\else the appendix.\fi
\end{enumerate}

\noindent They define non-interleaving as follows (paraphrased):
\begin{quote}
Suppose two sets of list elements $X$ and $Y$ satisfy:
\begin{itemize}
  \item All elements in $X$ were inserted concurrently to all elements in $Y$.
  \item The elements were inserted at the same location in the document, that is: after applying the insertions for $X \cup Y$ and their causal predecessors, $X \cup Y$ are contiguous in the list order.
\end{itemize}
Then either $X$ appears before $Y$ or vice-versa. That is, either $\forall x \in X, y \in Y.\, x \prec y$ or $\forall x \in X, y \in Y.\, y \prec x$, where $\prec$ is the order of elements in the final list.
\end{quote}

To show that no replicated list algorithm can satisfy this definition, it is sufficient to give a counterexample.
Starting from an empty list, suppose four replicas concurrently each insert one element.
After applying these four insertions, the list state must be some ordering of these four elements; let the order be $abcd$. Then $X = \{a, c\}$ and $Y = \{b, d\}$ satisfy the two hypotheses, but they are interleaved.
Since this situation could arise with any algorithm, it cannot be prevented.

In \Cref{sec:maximal} we give a new definition of non-interleaving that can be implemented, and we prove that our FugueMax algorithm implements it.

\section{The Fugue algorithm}
\label{sec:tree_fugue}

We now introduce \emph{Fugue} (pronounced \textipa{[fju:g]}), a new algorithm for replicated lists and collaborative text editing.
It is named after a form of classical music in which several melodic lines are interwoven in a pleasing way.
We analyze Fugue's non-interleaving properties in \Cref{sec:interleaving}, and we evaluate implementations in \Cref{sec:impl}. The FugueMax algorithm is a slight modification of Fugue which we defer until \Cref{sec:fugue_max_alg}.
\Cref{alg:tree_fugue} gives pseudocode for Fugue.


{
\auxfun{delete}
\auxfun{remove}
\auxfun{true}
\auxfun{false}
\auxfun{dom}

\begin{algorithm*}
  \caption{Pseudocode for the Fugue algorithm.}
  \label{alg:tree_fugue}
  \SetInd{0.5em}{0.5em}
  \DontPrintSemicolon
  \SetKwBlock{types}{types:}{}
  \SetKwBlock{cstate}{per-replica CRDT state:}{}
  \SetKwBlock{deliver}{on delivering}{}
  \SetKwBlock{kwfunction}{function}{}
  \SetKw{broadcast}{broadcast}

  \types{
    $\msf{RID}$, type of replica identifiers \;
    $\msf{ID} := (\msf{RID} \times \N) \cup \{\af{null}\}$, type of element IDs \;
    $\V$, type of values \;
    $\bot$, a marker for deleted nodes \;
    $\{L, R\}$, type of a child node's side (left or right)\;
    $\msf{NODE} := \msf{ID} \times (\V \cup \{\bot\}) \times \msf{ID} \times \{L, R\}$, tree node tuples $(\mathit{id}, \mathit{value}, \mathit{parent}, \mathit{side})$
  }
  \BlankLine

  \cstate{
    $\mathit{replicaID} \in \msf{RID}$: the unique ID of this replica \;
    $\mathit{tree} \subseteq \msf{NODE} \times \msf{ID}[] \times \msf{ID}[]$: a set of tree node triples $(\mathit{node}, \mathit{leftChildren}, \mathit{rightChildren})$, initially $\{(\mathit{root}, [], [])\}$ where $\text{\textit{root}} = (\af{null}, \bot, \af{null}, \af{null})$ \; 
    $\mathit{counter} \in \N$: a counter for generating element IDs, initially 0
  }
  \BlankLine

  \kwfunction({$\af{values}() : \V[]$}){
    \Return{$\af{traverse}(\msf{null})$}
  }
  \BlankLine

  \kwfunction({$\af{traverse}(\mathit{nodeID} : \msf{ID}) : \V[]$}){
      $\mathit{values} \gets []$\;
      $(\mathit{node}, \mathit{leftChildren}, \mathit{rightChildren}) \gets $ the unique triple $\in \mathit{tree}$ such that $node.id = \mathit{nodeID}$ \;
      \lFor {$\mathit{childId} \in \mathit{leftChildren}$} {
        $\mathit{values} \gets \mathit{values} + \af{traverse}(\mathit{childId})$
      }
      \lIf {$node.value \neq \bot$}{
        $\mathit{values} \gets \mathit{values} + [\mathit{node.value}]$
      }
      \lFor {$\mathit{childId} \in \mathit{rightChildren}$} {
        $\mathit{values} \gets \mathit{values} + \af{traverse}(\mathit{childId})$
      }
      \Return{$\mathit{values}$}
  }
  \BlankLine

  \kwfunction({$\af{insert}(i:\N, x:\V)$}){
    $\mathit{id} \gets (\mathit{replicaID}, \mathit{counter})$; $\mathit{counter} \gets \mathit{counter} + 1$ \;
    $\mathit{leftOrigin} \gets$ node for $(i-1)$-th value in $\af{values}()$, or $\mathit{root}$ if $i = 0$ \;
    $\mathit{rightOrigin} \gets$ next node after $\mathit{leftOrigin}$ in the tree traversal that includes tombstones \;
    \If {$\nexists\mathit{id}', v'.\,\, (\mathit{id}', v', \mathit{leftOrigin}.\mathit{id}, R) \in \mathit{tree}$} {
      $\mathit{node} \gets (\mathit{id}, x, \mathit{leftOrigin.id}, R)$ \label{line:gen_r_child} \quad// right child of $\mathit{leftOrigin}$; see \Cref{fig:insertions}(a)
    }
    \Else {
      $\mathit{node} \gets (\mathit{id}, x, \mathit{rightOrigin.id}, L)$ \quad// left child of $\mathit{rightOrigin}$; see \Cref{fig:insertions}(b)
    }
    \broadcast{$(\af{insert}, node)$} by causal broadcast
  }
  \BlankLine

  \deliver({$(\af{insert}, \mathit{node})$} by causal broadcast){
    $(\mathit{parent}, \mathit{leftSibs}, \mathit{rightSibs}) \gets $ the unique triple $\in \mathit{tree}$ such that $\mathit{parent.id} = \mathit{node}.\mathit{parent}$ \;
    \If{$\mathit{node.side} = R$} {
      $i \gets $ least index such that $\mathit{node.id} < \mathit{rightSibs}[i]$ \label{line:effect_r_child} \;
      insert $\mathit{node.id}$ into $\mathit{rightSibs}$ at index $i$
    } \Else {
      $i \gets $ least index such that $\mathit{node.id} < \mathit{leftSibs}[i]$ \;
      insert $\mathit{node.id}$ into $\mathit{leftSibs}$ at index $i$
    }
    $\mathit{tree} \gets \mathit{tree} \cup \{(\mathit{node}, [], [])\}$ \;
  }
  \BlankLine

  \kwfunction({$\delete(i:\N)$}){
    $\mathit{node} \gets $ node for $i$-th value in $\af{values}()$ \;
    \broadcast{$(\delete, node.id)$} by causal broadcast
  }
  \BlankLine

  \deliver({$(\delete, id)$} by causal broadcast){
    $(\mathit{node}, \textunderscore, \textunderscore) \gets $ the unique triple $\in \mathit{tree}$ such that $node.id = id$ \;
    $node.value \gets \bot$ \;
  }
  \BlankLine
\end{algorithm*}
}

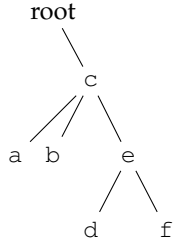
\begin{figure}
\centering
\begin{tikzpicture}
    \node (r) at (1.5,4) {root};
    \tikzstyle{every node}=[anchor=base,font=\ttfamily]
    \node (c) at (2.0,3) {c};
    \node (a) at (1.0,2) {a};
    \node (b) at (1.5,2) {b};
    \node (e) at (2.5,2) {e};
    \node (d) at (2.0,1) {d};
    \node (f) at (3.0,1) {f};
    \draw (r) -- (c);
    \draw (c) -- (a);
    \draw (c) -- (b);
    \draw (c) -- (e);
    \draw (e) -- (d);
    \draw (e) -- (f);
\end{tikzpicture}
\caption{One possible Fugue structure for the list \texttt{abcdef}. Observe that \texttt{a} and \texttt{b} are both left children of \texttt{c}; they are sorted lexicographically by their elements' IDs.}
\label{fig:binary_tree}
\end{figure}

We describe Fugue as an operation-based CRDT~\cite{Shapiro:2011}, although it can easily be reformulated as a state-based CRDT.
The external interface of Fugue is an ordered sequence of values, e.g., the characters in a text document.
Since the same value may appear multiple times in a list, we use \emph{element} to refer to a unique instance of a value.
Then the operations on the list are:
\begin{itemize}
  \item $\msf{insert}(i, x)$ (\Cref{alg:tree_fugue}, lines 21--38): Inserts a new element with value $x$ at index $i$, between existing elements at indices $i-1$ and $i$. All later elements (index $\ge i$) shift to an incremented index.
  \item $\msf{delete}(i)$ (\Cref{alg:tree_fugue}, lines 39--44): Deletes the element at index $i$. All later elements (index $\ge i + 1$) shift to a decremented index.
\end{itemize}
Note that we omit operations to mutate or move elements; these can be implemented by combining a replicated list with other CRDTs \cite{list_with_move}.
We also omit optimizations that compress consecutive runs of insertions or deletions; these can be added later without affecting the core algorithm.
At a high level the algorithm works as follows:

\paragraph*{\textbf{State}}
The state of each replica is a tree in which each non-root node is labeled with a unique ID and a value (\Cref{alg:tree_fugue}, line 10).
Each non-root node is marked as either a \emph{left} or \emph{right} child of its parent, but the tree is not necessarily binary: a parent can have multiple left children or right children, as illustrated in \Cref{fig:binary_tree}.
The tree does not need to be balanced.

Each non-root node in the tree corresponds to an element in the list (e.g., a character in the text document).
The list order is given by the depth-first in-order traversal over this tree: first recursively traverse a node's left children, then visit the node's own value, then traverse its right children (\Cref{alg:tree_fugue}, lines 12--20).
\emph{Same-side siblings}---nodes with the same parent and the same side---are traversed in lexicographic order of their IDs; the exact construction of IDs and their order is not important.

\begin{figure}
\centering
\begin{tikzpicture}
    \node at (0,4) {(a)};
    \node at (5,4) {(b)};
    \node (r1) at (1.5,4) {root};
    \node (r2) at (6.5,4) {root};
    \tikzstyle{every node}=[anchor=base,font=\ttfamily]
    \node (c1) at (2.0,3) {c};
    \node (a1) at (0.5,2) {a};
    \node (g1) at (1.0,1) [draw,circle] {g};
    \node (h1) at (0.5,0) {~};
    \node (b1) at (1.5,2) {b};
    \node (e1) at (2.5,2) {e};
    \node (d1) at (2.0,1) {d};
    \node (f1) at (3.0,1) {f};
    \draw (r1) -- (c1);
    \draw (c1) -- (a1);
    \draw [dashed] (a1) -- (g1);
    \draw (c1) -- (b1);
    \draw (c1) -- (e1);
    \draw (e1) -- (d1);
    \draw (e1) -- (f1);
    \node (c2) at (7.0,3) {c};
    \node (a2) at (5.5,2) {a};
    \node (g2) at (6.0,1) {g};
    \node (h2) at (5.5,0) [draw,circle] {h};
    \node (b2) at (6.5,2) {b};
    \node (e2) at (7.5,2) {e};
    \node (d2) at (7.0,1) {d};
    \node (f2) at (8.0,1) {f};
    \draw (r2) -- (c2);
    \draw (c2) -- (a2);
    \draw (a2) -- (g2);
    \draw [dashed] (g2) -- (h2);
    \draw (c2) -- (b2);
    \draw (c2) -- (e2);
    \draw (e2) -- (d2);
    \draw (e2) -- (f2);
\end{tikzpicture}
\caption{(a) Inserting a new element \texttt{g} between \texttt{a} and \texttt{b}. When \texttt{a} has no right children, making \texttt{g} a right child of \texttt{a} places it immediately after \texttt{a} in the traversal. (b) Inserting a second new element \texttt{h} between \texttt{a} and \texttt{g}. When \texttt{g} is a descendant of \texttt{a}, we make \texttt{h} a left child of \texttt{g}.}
\label{fig:insertions}
\end{figure}
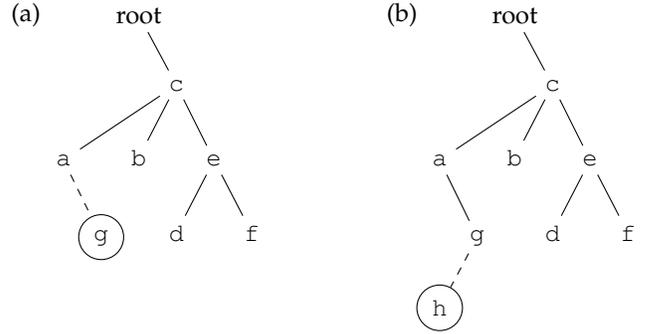

\paragraph*{\textbf{Insert}}
To implement $\msf{insert}(i, x)$, a replica creates a new node, labeled with a new unique ID and value $x$, at an appropriate position in its local tree: if the element at index $i-1$ has no right children, the new node becomes a right child of the element at index $i-1$ (lines 25--26); otherwise, the new node is added as a left child of the next element in the traversal order (lines 27--28).
\Cref{fig:insertions} illustrates how this choice is made, and \Cref{thm:correctness} shows that this approach results in the desired behavior.
The replica then uses a causal broadcast protocol~\cite{Birman:1991,Cachin:2011} to send the new node, its parent, and its side (left or right child) to other replicas (line 29), which add the node to their own local trees (lines 30--38).
We assume the causal broadcast protocol immediately delivers the message to the sender without waiting for any network communication.

A replica will not create a new node where it already has a same-side sibling, i.e., it will try to keep the tree binary. However, multiple replicas may concurrently insert nodes at the same position, creating same-side siblings like \texttt{a} and \texttt{b} in \Cref{fig:binary_tree}.

\paragraph*{\textbf{Delete}}
To implement $\msf{delete}(i)$, a replica looks up the node at index $i$ in the current list state, then causally broadcasts a message containing that element's ID (lines 40--41).
All replicas then replace that node's value with a special value $\bot$ (line 44), flagging it as deleted (i.e., making it a \emph{tombstone}).
Nodes with this value are skipped when computing the externally visible list order and indexes (line 18); however, their non-deleted descendants are still traversed normally, and a deleted node may still be used as a parent of a new node.

We cannot remove a deleted element's node entirely: it may be an ancestor to non-deleted nodes, including nodes inserted concurrently.
In \Cref{sec:impl} we discuss ways of mitigating memory usage from tombstones.


\begin{theorem}
\label{thm:correctness}
\Cref{alg:tree_fugue} satisfies the strong list specification of Attiya et al.~\cite{attiya}.
\end{theorem}
\begin{IEEEproof}
For any execution, we must show that there is a total order $\prec$ over all list elements (across all replicas), such that:
\begin{enumerate}[(a)]
  \item At any time, calling $\msf{values}()$ on a replica returns the list of values corresponding to all elements for which the replica received $\msf{insert}$ messages, minus the elements for which it received $\msf{delete}$ messages, in order $\prec$.
  \item Suppose a replica's $\msf{values}()$ query yields values corresponding to elements $[a_0, a_1, \dots, a_{n-1}]$ just before the insert generator $\msf{insert}(i, x)$ is called. Then the inserted element $e$ satisfies $a_0, a_1, \dots, a_{i-1} \prec e \prec a_i, a_{i+1}, \dots, a_{n-1}$.
\end{enumerate}
Let $\prec$ be the total order given by the depth-first in-order traversal on the union of all replicas' local trees (with tombstone nodes overriding nodes with the originally inserted value). To show (a), note that by the causal order delivery assumption, a $\msf{delete}$ message is received after its corresponding $\msf{insert}$ message. Therefore, on any given replica, the set of tree nodes with $\mathit{value} \neq \bot$ are those nodes that have been inserted but not deleted on that replica. These are exactly the nodes whose values are returned by $\msf{values}()$, in the same order as $\prec$ because the same traversal is used.

To show (b), note that $\mathit{leftOrigin}$ and $\mathit{rightOrigin}$ are consecutive elements in the tree traversal, and $\mathit{leftOrigin} = a_{i-1}$, the non-tombstone node immediately preceding the insertion position.
If $\mathit{leftOrigin}$ has no right children, inserting the new node as a right child of $\mathit{leftOrigin}$ makes the new node the immediate successor of $\mathit{leftOrigin}$ in the tree traversal.
If $\mathit{leftOrigin}$ does have right children, $\mathit{rightOrigin}$ must be a descendant of $\mathit{leftOrigin}$, and $\mathit{rightOrigin}$ must have no left children (since otherwise $\mathit{leftOrigin}$ and $\mathit{rightOrigin}$ would not be consecutive), and therefore inserting the new node as a left child of $\mathit{rightOrigin}$ ensures the traversal visits the new child between $\mathit{leftOrigin}$ and $\mathit{rightOrigin}$.
In either case, the newly inserted element appears between $a_{i-1}$ and $a_i$ in the traversal, as required.
\end{IEEEproof}

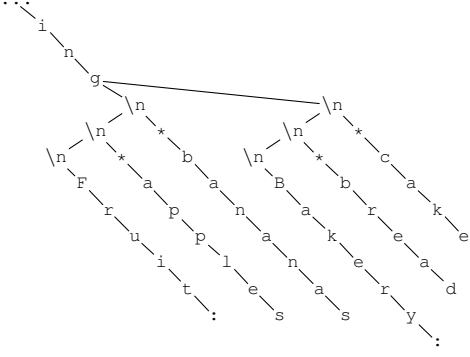
\begin{figure}
\centering
\newcommand{\fuguechars}[4]{%
    \foreach \i/\letter in {#4} {
        \node (#1\i) at (#2+\i,#3-\i) {\letter};
        \ifnum\i>0
            \pgfmathsetmacro\prev{int(\i-1)};
            \draw (#1\prev) -- (#1\i);
        \fi
    }
}%
\begin{tikzpicture}[scale=0.35]
    \tikzstyle{every node}=[anchor=base,font=\ttfamily\scriptsize,inner sep=0.5pt]
    \fuguechars{shopping}{1}{12}{0/\dots, 1/i, 2/n, 3/g}
    \fuguechars{bananas}{5.5}{8}{0/\textbackslash{}n, 1/*, 2/b, 3/a, 4/n, 5/a, 6/n, 7/a, 8/s}
    \fuguechars{apples}{4}{7}{0/\textbackslash{}n, 1/*, 2/a, 3/p, 4/p, 5/l, 6/e, 7/s}
    \fuguechars{fruit}{2.5}{6}{0/\textbackslash{}n, 1/F, 2/r, 3/u, 4/i, 5/t, 6/:}
    \fuguechars{cake}{13}{8}{0/\textbackslash{}n, 1/*, 2/c, 3/a, 4/k, 5/e}
    \fuguechars{bread}{11.5}{7}{0/\textbackslash{}n, 1/*, 2/b, 3/r, 4/e, 5/a, 6/d}
    \fuguechars{bakery}{10}{6}{0/\textbackslash{}n, 1/B, 2/a, 3/k, 4/e, 5/r, 6/y, 7/:}
    \draw (shopping3) -- (bananas0);
    \draw (shopping3) -- (cake0);
    \draw (bananas0) -- (apples0) -- (fruit0);
    \draw (cake0) -- (bread0) -- (bakery0);
\end{tikzpicture}
\caption{Fugue tree after executing the editing history in \Cref{fig:backward-interleaving}. Since the two users' insertions are in separate subtrees, a depth-first traversal does not interleave them.}
\label{fig:shopping_tree}
\end{figure}

\section{Formalizing Non-Interleaving}
\label{sec:interleaving}
We have proven that Fugue satisfies the strong list specification (\Cref{thm:correctness}). We now show that it also avoids the interleaving problem described in \Cref{sec:related} except in specific, rare situations where some interleaving is inevitable.

Specifically, we show that a variant of Fugue, FugueMax, is \emph{maximally non-interleaving}: FugueMax avoids interleaving of both forward and backward insertions, to a maximum possible extent. We then show that Fugue differs from FugueMax only rarely while allowing a simpler algorithm. In practice, Fugue's simplicity likely outweighs FugueMax's slightly better non-interleaving guarantees.

Intuitively, non-interleaving holds because concurrent edits end up in different subtrees, which are traversed separately.
This is illustrated in \Cref{fig:shopping_tree}, which shows the Fugue representation of the editing history in \Cref{fig:backward-interleaving}.






\subsection{Preliminary Definitions}

In an execution using a replicated list, the \emph{left origin} of an element is the element directly preceding its insertion position at the time of insertion. Specifically, if the element was inserted by an $\msf{insert}(i, x)$ call, then its left origin was at index $i-1$ at the time of this call. If there was no such element ($i=0$), then its left origin is a special symbol $\mathit{start}$. This definition coincides with the $\mathit{leftOrigin}$ variable in \Cref{alg:tree_fugue}, except using $\mathit{start}$ instead of $\mathit{root}$. We sometimes denote the left origin of $A$ by $A.\mathit{leftOrigin}$.

Dually to the definition of left origin, we define the \emph{right origin} of an element to be the element directly following its insertion position at the time of insertion, or the special symbol $\mathit{end}$ if no following element exists. Specifically, the right origin is the element directly following the left origin in \emph{the list including tombstones (deleted elements)}, like the $\mathit{rightOrigin}$ variable in \Cref{alg:tree_fugue}. This choice simplifies the analysis by letting us ignore deletions.

Define the \emph{left-origin tree} to be the tree of list elements in which each element's parent is its left origin. Observe that the tree is rooted at $\mathit{start}$, because every element was inserted either at the beginning of the list, or after some existing element (i.e.\ an existing tree node). This tree's definition is similar to causal trees \cite{grishchenko} and timestamped insertion trees \cite{attiya}.

Similarly, define the \emph{right-origin tree} to be the tree of list elements in which each element's parent is its right origin. This tree is rooted at $\mathit{end}$. For a somewhat complicated example of both trees, see \Cref{fig:maximal-trees}.

If list element $A$ had already been inserted at the time when list element $B$ was inserted, we say that $A$ is \emph{causally prior} to $B$, and $B$ is \emph{causally later} than $A$ (regardless of whether $A$ or $B$ were subsequently deleted again).
In particular, the left and right origins of a list element are always causally prior to that element.
When $A$ is neither causally prior nor causally later than $B$, we say that $A$ and $B$ are \emph{concurrent}.
This relation defines a partial order over list elements, and the causal broadcast protocol in \Cref{alg:tree_fugue} delivers the insertion operations for those list elements in a linear extension of this partial order.

\subsection{Defining Maximal Non-Interleaving}
\label{sec:maximal}

We already saw that the definition of non-interleaving by Kleppmann et al.~\cite{interleaving} is impossible to satisfy (\Cref{sec:papoc-paper}). Finding a satisfiable definition requires some care.
Let us first consider non-interleaving for forward insertions, like those in \Cref{fig:forward-interleaving}.

\begin{definition}
\label{def:forward}
A replicated list algorithm is \emph{forward non-interleaving} if it satisfies the strong list specification~\cite{attiya} and the following holds for all list elements $A$ and $B$ in all possible list states: if $A$ is the left origin of $B$, and $B$ appears earlier in the list than any other element that has $A$ as left origin, then $A$ and $B$ are consecutive list elements.
\end{definition}

This definition captures the idea that elements $A,B$ inserted in a forward sequence should be consecutive---no other elements may interleave with the sequence $AB$. In particular, this holds when $B$ is the \emph{only} element that has $A$ as left origin. Once multiple elements $B,C$ have left origin $A$, we cannot place all of them immediately after $A$. But we can still guarantee that sequences inserted after $B$ and $C$ respectively are not interleaved:

\begin{lemma}
Let a replicated list algorithm satisfy forward non-interleaving as defined in \Cref{def:forward}. Then it is also satisfies Kleppmann et al.'s definition of forward non-interleaving~\cite[\S 4]{opsets}: if two sequences $B_1...B_m$ and $C_1...C_n$ are inserted from left to right concurrently at the same starting position, then all $B_j$ are on the same side of all $C_i$ in the final list order.
\end{lemma}
\begin{IEEEproof}
Let $\prec$ be the global total order on list elements guaranteed by the strong list specification. Without loss of generality, $B_1 \prec C_1$. We need to show that $B_2, \dots, B_m \prec C_1$.

It is possible for a replica to be in a state that contains $C_1$, $B_2$, and all causally prior elements (including $B_1$), but no others. Since $C_1$ was inserted concurrently to $B_1$, $B_2$ is the only element in this state that is causally later than $B_1$. Thus it is the only element with left origin $B_1$. Then by forward non-interleaving, $B_1$ and $B_2$ are consecutive. Together with $B_1 \prec C_1$, this implies $B_2 \prec C_1$.
Because this relation holds for one possible replica in one state, the strong list specification requires it to hold in the final list order.

A similar argument shows that $B_3, \dots, B_m \prec C_1$.
\end{IEEEproof}

It is tempting to define ``backward non-interleaving'' analogously to forward non-interleaving (replacing left origins with right origins), then define ``non-interleaving'' as the conjunction of forward and backward non-interleaving. 

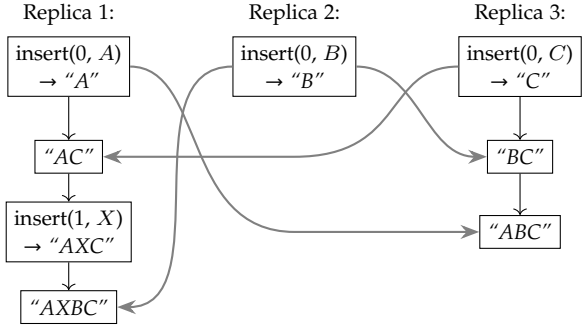
\begin{figure}
    \centering
    \begin{tikzpicture}[font=\footnotesize]
        \tikzstyle{effector}=[draw,rectangle]
        \tikzstyle{generator}=[matrix,draw,rectangle,inner sep=1.5pt]
        \tikzstyle{network}=[thick,gray,-{Stealth[length=2.5mm]}]
        \node at (0,7.9) {Replica 1:};
        \node at (3,7.9) {Replica 2:};
        \node at (6,7.9) {Replica 3:};
        \node (left1) at (0,7.2) [generator] {
            \node {insert(0, $A$)};\\
            \node {$\rightarrow$ \emph{``A''}};\\
        };
        \node (mid1) at (3,7.2) [generator] {
            \node {insert(0, $B$)};\\
            \node {$\rightarrow$ \emph{``B''}};\\
        };
        \node (right1) at (6,7.2) [generator] {
            \node {insert(0, $C$)};\\
            \node {$\rightarrow$ \emph{``C''}};\\
        };
        \node (left2) at (0,6) [effector] {\emph{``AC''}};
        \node (right2) at (6,6) [effector] {\emph{``BC''}};
        \node (right3) at (6,5) [effector] {\emph{``ABC''}};
        \node (left3) at (0,5) [generator] {
            \node {insert(1, $X$)};\\
            \node {$\rightarrow$ \emph{``AXC''}};\\
        };
        \node (left4) at (0,4) [effector] {\emph{``AXBC''}};
        \draw [->] (left1) -- (left2);
        \draw [->] (left2) -- (left3);
        \draw [->] (left3) -- (left4);
        \draw [->] (right1) -- (right2);
        \draw [->] (right2) -- (right3);
        \draw [network] (left1.east) .. controls +(right:1cm) and +(left:1.5cm) .. (3,5) .. controls +(right:1.5cm) and +(left:1cm) .. (right3.west);
        \draw [network] (mid1.east) to [out=0,in=180] (right2.west);
        \draw [network] (mid1.west) to [out=180,in=0] (left4.east);
        \draw [network] (right1.west) .. controls +(left:1cm) and +(right:1.5cm) .. (3,6) .. controls +(left:1.5cm) and +(right:1cm) .. (left2.east);
    \end{tikzpicture}
    \caption{An execution in which it is impossible to achieve both forward and backward non-interleaving, but maximal non-interleaving is possible.}
    \label{fig:force-backward}
\end{figure}

However, there are exceptional executions in which forward non-interleaving \emph{forces} us to interleave backward insertions. \Cref{fig:force-backward} gives an example: Starting from an empty list, three replicas concurrently insert $A$, $B$, and $C$. Replica 3 receives all three elements and puts them in some order; without loss of generality, it is $A \prec B \prec C$. Replica 1 receives $A$ and $C$, then inserts $X$ in between those elements to obtain $AXC$. Finally, Replica 1 receives $B$.

Since $X$ is the only element with left origin $A$, forward non-interleaving requires them to be consecutive: \emph{AX}. Also, the strong list specification requires that since $A \prec B \prec C$ on Replica 3, all other replicas must place those elements in the same order. Thus the final list order must be \emph{AXBC}. But then $X$ and $C$ are not consecutive, even though $X$ is the only element with right-origin $C$. This rules out the above version of backward non-interleaving.

We do still want to mandate backward non-interleaving whenever it is possible. When forward and backward non-interleaving are in conflict with each other, we let forward non-interleaving take precedence, since forward (left-to-right) insertions are more common in text editors. 

\begin{definition}
\label{def:maximal}
A replicated list algorithm is \emph{maximally non-interleaving} if it satisfies the strong list specification~\cite{attiya} and the following holds for all list elements $A$ and $B$ in all possible list states:
\begin{enumerate}[(1)]
    \item (Forward non-interleaving) If $A$ is the left origin of $B$, and $B$ appears earlier in the list than any other element that has $A$ as left origin, then $A$ and $B$ are consecutive list elements.\label{cond:1}
    \item (Backward non-interleaving, with exceptions) If $B$ is the right origin of $A$, and $A$ appears later in the list than any other element that has $B$ as right origin, then $A$ and $B$ are consecutive list elements, \emph{unless} \Cref{lem:2prime} below says otherwise.\label{cond:2}
    \item If $A$ and $B$ have the same left origin and the same right origin, then the element with the lower ID appears earlier in the list.\label{cond:3}
\end{enumerate}
\end{definition}

\begin{lemma}
\label{lem:2prime}
Given a replicated list algorithm that satisfies the strong list specification and forward non-interleaving (condition (1)). Suppose $B$ is the right origin of $A$, and $A$ appears later in the list than any other element that has $B$ as right origin, \emph{but}:
\begin{enumerate}[i.]
    \item $A$ and $B$ have different left origins; and
    \item there is a $C$ in the current list state such that $A.\mathit{leftOrigin} \prec C \prec B$ and $C$ is not a descendant of $A.\mathit{leftOrigin}$ in the left-origin tree.
\end{enumerate}
Then $A \prec C \prec B$, so $A$ and $B$ are not consecutive list elements.
\end{lemma}

We will prove \Cref{lem:2prime} in the next section. For now, observe that the lemma forces $X \prec B \prec C$ in the example of \Cref{fig:force-backward}. Thus the list order \emph{AXBC} is allowed by maximal non-interleaving: the fact that $X$ and $C$ are not consecutive is permitted due to the exception in condition (2).

Condition (3) is an arbitrary choice; two elements with the same left origin and the same right origin were inserted concurrently at the exact same place, so there is no reason to order them in a particular way.
It turns out that this arbitrary choice is the \emph{only} remaining degree of freedom after assuming (1) and (2): we show in \Cref{sec:maximal_unique} that maximal non-interleaving uniquely determines the list order. 

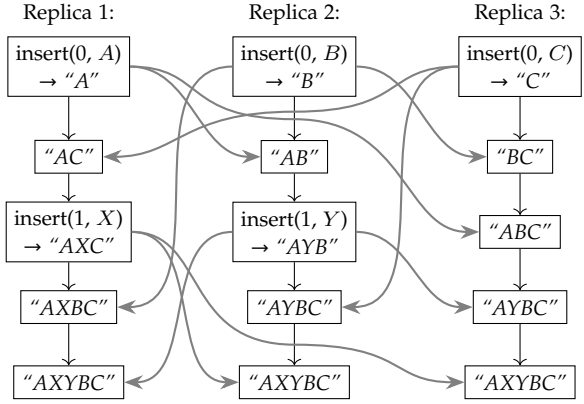
\begin{figure}
    \centering
    \begin{tikzpicture}[font=\footnotesize]
        \tikzstyle{effector}=[draw,rectangle]
        \tikzstyle{generator}=[matrix,draw,rectangle,inner sep=1.5pt]
        \tikzstyle{network}=[thick,gray,-{Stealth[length=2.5mm]}]
        \node at (0,7.9) {Replica 1:};
        \node at (3,7.9) {Replica 2:};
        \node at (6,7.9) {Replica 3:};
        \node (left1) at (0,7.2) [generator] {
            \node {insert(0, $A$)};\\
            \node {$\rightarrow$ \emph{``A''}};\\
        };
        \node (mid1) at (3,7.2) [generator] {
            \node {insert(0, $B$)};\\
            \node {$\rightarrow$ \emph{``B''}};\\
        };
        \node (right1) at (6,7.2) [generator] {
            \node {insert(0, $C$)};\\
            \node {$\rightarrow$ \emph{``C''}};\\
        };
        \node (left2) at (0,6) [effector] {\emph{``AC''}};
        \node (mid2) at (3,6) [effector] {\emph{``AB''}};
        \node (right2) at (6,6) [effector] {\emph{``BC''}};
        \node (right3) at (6,5) [effector] {\emph{``ABC''}};
        \node (left3) at (0,5) [generator] {
            \node {insert(1, $X$)};\\
            \node {$\rightarrow$ \emph{``AXC''}};\\
        };
        \node (mid3) at (3,5) [generator] {
            \node {insert(1, $Y$)};\\
            \node {$\rightarrow$ \emph{``AYB''}};\\
        };
        \node (left4) at (0,4) [effector] {\emph{``AXBC''}};
        \node (left5) at (0,3) [effector] {\emph{``AXYBC''}};
        \node (mid4) at (3,4) [effector] {\emph{``AYBC''}};
        \node (mid5) at (3,3) [effector] {\emph{``AXYBC''}};
        \node (right4) at (6,4) [effector] {\emph{``AYBC''}};
        \node (right5) at (6,3) [effector] {\emph{``AXYBC''}};
        \draw [->] (left1) -- (left2);
        \draw [->] (left2) -- (left3);
        \draw [->] (left3) -- (left4);
        \draw [->] (left4) -- (left5);
        \draw [->] (mid1) -- (mid2);
        \draw [->] (mid2) -- (mid3);
        \draw [->] (mid3) -- (mid4);
        \draw [->] (mid4) -- (mid5);
        \draw [->] (right1) -- (right2);
        \draw [->] (right2) -- (right3);
        \draw [->] (right3) -- (right4);
        \draw [->] (right4) -- (right5);
        \draw [network] (left1.east) to [out=0,in=180] (mid2.west);
        \draw [network] (left1.east) .. controls +(right:1cm) and +(left:1.5cm) .. (3,6.5) .. controls +(right:1.5cm) and +(left:1cm) .. (right3.west);
        \draw [network] (left3.east) to [out=0,in=180] (mid5.west);
        \draw [network] (left3.east) .. controls +(right:1cm) and +(left:1.5cm) .. (3,3.5) .. controls +(right:1.5cm) and +(left:1cm) .. (right5.west);
        \draw [network] (mid1.east) to [out=0,in=180] (right2.west);
        \draw [network] (mid1.west) to [out=180,in=0] (left4.east);
        \draw [network] (mid3.west) to [out=180,in=0] (left5.east);
        \draw [network] (mid3.east) to [out=0,in=180] (right4.west);
        \draw [network] (right1.west) .. controls +(left:1cm) and +(right:1.5cm) .. (3,6.6) .. controls +(left:1.5cm) and +(right:1cm) .. (left2.east);
        \draw [network] (right1.west) to [out=180,in=0] (mid4.east);
    \end{tikzpicture}
    \caption{An extension of the execution in \Cref{fig:force-backward}, demonstrating the conditions of maximal non-interleaving.}
    \label{fig:maximal-noninterleaving}
\end{figure}

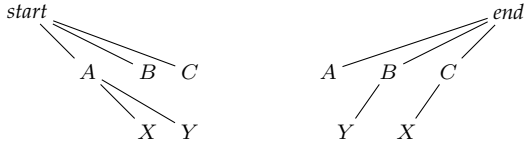
\begin{figure}
    \centering
    \begin{tikzpicture}[font=\footnotesize,scale=0.8,inner sep=2pt]
        \node (start) at (0,4) {\emph{start}};
        \node (a1) at (1,3) {$A$};
        \node (b1) at (2,3) {$B$};
        \node (c1) at (2.7,3) {$C$};
        \node (x1) at (2,2) {$X$};
        \node (y1) at (2.7,2) {$Y$};
        \draw (start) -- (a1);
        \draw (start) -- (b1);
        \draw (start) -- (c1);
        \draw (a1) -- (x1);
        \draw (a1) -- (y1);
        \node (end) at (8,4) {\emph{end}};
        \node (a2) at (5,3) {$A$};
        \node (b2) at (6,3) {$B$};
        \node (c2) at (7,3) {$C$};
        \node (y2) at (5.3,2) {$Y$};
        \node (x2) at (6.3,2) {$X$};
        \draw (end) -- (a2);
        \draw (end) -- (b2);
        \draw (end) -- (c2);
        \draw (b2) -- (y2);
        \draw (c2) -- (x2);
    \end{tikzpicture}
    \caption{Left- and right-origin trees for \Cref{fig:maximal-noninterleaving}. The Fugue tree in this example is equivalent to the left-origin tree (all elements are right children of their parent).}
    \label{fig:maximal-trees}
\end{figure}

\Cref{fig:maximal-noninterleaving} shows how maximal non-interleaving determines the list order in a more complex execution; see \Cref{fig:maximal-trees} for the final left- and right-origin trees.
We start as in \Cref{fig:force-backward} with three replicas inserting $A$, $B$, and $C$ concurrently into an empty list.
These three elements have the same left and right origins (\emph{start} and \emph{end} respectively), so condition (3) requires them to be in ID order; without loss of generality, assume $A \prec B \prec C$.
Replica 1 then receives $\{A,C\}$ and inserts $X$ between them, while concurrently Replica 2 receives $\{A,B\}$ and inserts $Y$ between them.
We have already seen in \Cref{fig:force-backward} that in the state \emph{AXBC} on Replica 1, condition (1) requires \emph{AX} to be consecutive, and thus $X \prec B$ in the list order.
Moreover, $Y$ is the only element with right origin $B$, so condition (2) requires \emph{YB} to be consecutive (nothing contradicts this requirement).
Finally, $X$ is the only element with right origin $C$, but since condition \Cref{lem:2prime} requires $X \prec B \prec C$, it is an exception from condition (2), and thus \emph{XC} is not required to be consecutive.
Thus, the only list order that satisfies maximal non-interleaving in this example is \emph{AXYBC}.

\subsection{From Fugue to FugueMax}
\label{sec:fugue_max_alg}
It turns out that in executions like \Cref{fig:maximal-noninterleaving}, Fugue might \emph{not} satisfy maximal non-interleaving. Indeed, the previous paragraph explained that maximal non-interleaving implies $X \prec Y$. But in the Fugue tree (\Cref{fig:maximal-trees}'s left side), $X$ and $Y$ are same-side siblings, hence traversed in the lexicographic order of their IDs. This order might be $Y \prec X$.

We can repair this issue by changing the order of right-side siblings in the Fugue tree: when siblings $X$ and $Y$ have different right origins $C \neq B$, we sort $X$ and $Y$ by the \emph{reverse} order of their right origins. For example, in \Cref{fig:maximal-trees}'s Fugue tree, $B \prec C$, so we should order $X \prec Y$. That allows $Y$ to be consecutive with $B$, the leftmost right origin.
We learned of this reverse-right-origin technique from Gentle's YjsMod \cite{gentle_code}.

\begin{definition}
\emph{FugueMax} is the replicated list algorithm that is identical to Fugue except that its tree traversal visits right-side siblings in the \emph{reverse} order of their right origins, breaking ties using the lexicographic order of their IDs.
\end{definition}

Concretely, FugueMax's algorithm is identical to Fugue's \Cref{alg:tree_fugue} except:
\begin{enumerate}[1.]
    \item When generating a right child (Line~\ref{line:gen_r_child}), the node is additionally tagged with its right origin: $\mathit{node} \gets (\mathit{id}, x, \mathit{leftOrigin.id}, R, \mathbf{\mathit{rightOrigin.id}})$.
    \item On delivering a right child (Line~\ref{line:effect_r_child}), it uses the sibling order described above: $i \gets $ the least index such that $\mathit{node.rightOrigin} \succ \mathit{rightSibs}[i].\mathit{rightOrigin}$ \textbf{or} $\big(\mathit{node.rightOrigin} = $ $\mathit{rightSibs}[i].\mathit{rightOrigin}$ \textbf{and} $\mathit{node.id} < \mathit{rightSibs}[i]\big)$, where $\succ$ is the existing list order and $<$ is the lexicographic order on node IDs.
\end{enumerate}
The proof that Fugue satisfies the strong list specification (\Cref{thm:correctness}) works identically for FugueMax.

We show in \Cref{sec:maximal_proof} that FugueMax is maximally non-interleaving and that Fugue is forward non-interleaving.

Moreover, one can check that in any situation where Fugue and FugueMax differ, some interleaving is inevitable, for any algorithm. Indeed, Fugue and FugueMax only differ in a situation analogous to \Cref{fig:maximal-noninterleaving}: there must exist elements with the same left origin ($X$ and $Y$) but different right origins ($B$ and $C$). In such a situation, forward non-interleaving necessitates $\{X, Y\} < \{B, C\}$, but backward non-interleaving only allows the incompatible orders $YBXC$ or $XCYB$. (Consider tree traversals in \Cref{fig:maximal-trees}.)

Such situations involve multiple interacting concurrent updates, which should be rare in practice. The advantage of FugueMax is only that it backward-interleaves one fewer pair of characters (here $YB$). Thus we believe that in practice, Fugue's simpler algorithm is worth its small amount of additional interleaving.

\subsection{FugueMax is Maximally Non-Interleaving}
\label{sec:maximal_proof}
The goal of this section is to prove that FugueMax is maximally non-interleaving (\Cref{thm:maximal_proof}).

Recall that a depth-first pre-order traversal of a tree follows the rule: visit a node, then traverse its children in some order (whereby all descendants of a child are visited recursively before moving on to the next child).

\begin{lemma}
\label{lem:rga_tree}
A replicated list algorithm that satisfies the strong list specification is forward non-interleaving if and only if its list order is some depth-first pre-order traversal of the left-origin tree.
\end{lemma}
\begin{IEEEproof}
$(\Longleftarrow)$: In any list state, let $A$ and $B$ be list elements such that $A$ is the left origin of $B$, and $B$ appears earlier in the list than any other element that has $A$ as left origin. In the left-origin tree, $B$ is the child of $A$ that appears earliest in the list. Thus the list's depth-first pre-order traversal will visit $B$ immediately after $A$: $A$ and $B$ are consecutive list elements.

$(\Longrightarrow)$: To show that the list order is a depth-first pre-order traversal of the left-origin tree, it suffices to prove:
\begin{enumerate}[(a)]
    \item If $A$ is the parent of $B$ in the left-origin tree, then $A \prec B$ in the list order.
    \item If $A$ and $D$ are siblings in the tree and $A \prec D$, then the entire subtree rooted at $A$ precedes $D$ in the list order.
\end{enumerate}

For statement (a), $A$ is the left origin of $B$, so $A \prec B$ by the strong list specification.

For statement (b), first observe that $D$ is not causally later than $A$. We show this by contradiction: assume $D$ is inserted in a list state that already contains $A$. Since $A$ and $D$ are siblings in the left-origin tree, $D.\mathit{leftOrigin} = A.\mathit{leftOrigin}$, and $A.\mathit{leftOrigin} \prec A$ in the list order. At the time $D$ is inserted, $D.\mathit{leftOrigin}$ and $D$ must be consecutive, implying $D \prec A$, which contradicts $A \prec D$.

Next, let $B$ be a child of $A$ in the left-origin tree. It is possible for a replica to be in a state that contains $B$, $D$, and all causally prior elements (including $A$), but no others. Every element $E \notin \{B,D\}$ in this state is causally prior to either $B$ or $D$; however, if $E$ is causally later than $A$, then $E$ cannot be causally prior to $D$, since this would contradict the previous paragraph's finding that $A$ is not causally prior to $D$. Therefore, any element of this state that is causally later than $A$ and not equal to $B$ must be causally prior to $B$. Any element whose left origin is $A$ must be causally later than $A$. Hence $B$ appears earlier in the list than any other element whose left origin is $A$. By forward non-interleaving, $A$ and $B$ are consecutive in this state. Thus $B \prec D$.

One can similarly prove that $B \prec D$ whenever $B$ is a grandchild, great-grandchild, etc. of $A$, using induction on the depth of $B$. Hence all of the elements in the subtree rooted at $A$ appear before $D$ in the list order.
\end{IEEEproof}

\begin{IEEEproof}[{Proof of \Cref{lem:2prime}}]
By \Cref{lem:rga_tree}, the list order must be some depth-first pre-order traversal of the left-origin tree. In any such traversal, $C$ appears later in the list order than the entire subtree rooted at $A.\mathit{leftOrigin}$, because it appears later than $A.\mathit{leftOrigin}$ and is not part of its subtree. In particular, $A \prec C$. Meanwhile, $C \prec B$ by assumption.
\end{IEEEproof}

\begin{lemma}
\label{lem:fugue_left_origin}
In any state of FugueMax (resp., Fugue), we have the following facts about left origins.
\begin{enumerate}[(a)]
    \item The left origin of an element $D$ is given by: starting at $D$, walk up the FugueMax tree until you encounter a node that is a right child of its parent; that node's parent is $D$'s left origin.
    \item Let $A$ and $B$ be list elements such that $A \prec B$. Then $B$ is a descendant of $A$ in the FugueMax tree if and only if $B$ is a descendant of $A$ in the left-origin tree.
\end{enumerate}
\end{lemma}
\begin{IEEEproof}
\textbf{(a).}
Let $E$ be the alleged left origin for $D$. In the state just after inserting $D$, list elements $E$ and $D$ are consecutive in FugueMax's tree traversal. Since FugueMax satisfies the strong list specification (by the same proof as \Cref{thm:correctness}), $E$ must be $D$'s left origin.

\textbf{(b).} $(\Longleftarrow)$: Assume $B$ is a child of $A$ in the left-origin tree, that is, $B$ was immediately after $A$ in the list order when $B$ was inserted. Then the FugueMax algorithm either made $B$ a right child of $A$ in the FugueMax tree (if $A$ had no right child at the time of inserting $B$), or a left child of a right descendant of $A$ (if $A$ had at least one right child). In either case, $B$ is a descendant of $A$ in the FugueMax tree. By induction we generalize to the case when $B$ is a descendant of $A$ in the left-origin tree, using a similar argument at each tree level.

$(\Longrightarrow)$: Assume $B$ is a descendant of $A$ in the FugueMax tree and $A \prec B$. If $B$ is a left child of its parent, $B$ appears before its parent in the list order, so the parent cannot be $A$. Walking up the tree from $B$, since there is an element before $B$ in the list order, we must eventually reach an element that is a right child of its parent $P$. By (a), $P$ must be the left origin of $B$. Either $P=A$, in which case $B$ is a child of $A$ in the left-origin tree, or (by induction) $P$ is a descendant of $A$ in the left-origin tree.
\end{IEEEproof}

\begin{theorem}
\label{thm:maximal_proof}
FugueMax satisfies conditions (1), (2), and (3) of \Cref{def:maximal}. Hence FugueMax is maximally non-interleaving.
\end{theorem}
\begin{IEEEproof}
\textbf{Condition (1).}
Let $A$ and $B$ have the form in condition (1): $A$ is the left origin of $B$, and $B$ appears earlier in the list than any other element that has $A$ as left origin. We must prove that $A$ and $B$ are consecutive.

By \Cref{lem:fugue_left_origin}(a), in the FugueMax tree, $B$ is the earliest among all nodes that are either a right child of $A$ or a left descendant of a right child of $A$. This is precisely the first node that FugueMax's tree traversal visits when traversing the right children of $A$. Thus the traversal visits $B$ immediately after $A$: they are consecutive.

\textbf{Condition (2).}
Let $A$ and $B$ have the form in condition (2): $B$ is the right origin of $A$, $A$ appears later in the list than any other element that has $B$ as right origin, and conditions (i) and (ii) in \Cref{lem:2prime} are not both true. We must prove that $A$ and $B$ are consecutive.

First suppose (i) is false, so that $A$ and $B$ have the same left origins. By \Cref{lem:fugue_left_origin}(b), $B$ is a descendant of $A.\mathit{leftOrigin}$ in the FugueMax tree, so $A.\mathit{leftOrigin}$ must have at least one right child in the FugueMax tree. Thus, when $A$ was inserted between $A.\mathit{leftOrigin}$ and $B$, FugueMax's $\msf{insert}$ function made $A$ a left child of $B$.

Because $A$ is the last element in the list order with right origin $B$, $A$ is also the last to be traversed of the left children of $B$ in the FugueMax tree. Hence if $A$ has no right children of its own, then FugueMax's tree traversal visits $B$ immediately after $A$: they are consecutive. Otherwise, let $D$ be a right child of $A$ such that $A$ did not have any other right child when $D$ was inserted. Immediately before the insertion of $D$, $A$ did not have a right-side child, so $A$ and $B$ were consecutive. Hence $D$'s right origin is $B$ and $A \prec D$, contrary to assumption.

Next, suppose (i) is true and $A$ and $B$ are not consecutive. We must prove that there exists a $C$ of the form given in (ii).
Since $A$ and $B$ are not consecutive, there exist one or more elements in between $A$ and $B$; take $C$ to be an in-between element that is not causally later than any other element between $A$ and $B$.
We have $A.\mathit{leftOrigin} \prec C \prec B$ because $A.\mathit{leftOrigin} \prec A$. It remains to prove that $C$ is \emph{not} a descendant of $A.\mathit{leftOrigin}$ in the left-origin tree.

We use proof by contradiction: assume that $C$ is a descendant of $A.\mathit{leftOrigin}$ in the left-origin tree. Thus $C$ is either a descendant of $A$ or a sibling (it cannot be a descendant of a sibling, since otherwise the sibling would be a causally prior element between $A$ and $B$).

\textit{Case $C$ is a descendant of $A$ in the left-origin tree.}
In this case, $C$ is causally later than both $A$ and $B$ and was inserted in between them. Because $A$ appears later in the list than any other element that has $B$ as right origin, $C$'s right origin is not $B$; instead, it must be some other element between $A$ and $B$. But then $C$'s right origin is an in-between element that is causally prior to $C$, contradicting our choice of $C$.

\textit{Case $C$ is a sibling of $A$ in the left-origin tree.}
That is, $A$ and $C$ have the same left origins.

At the time of $A$'s insertion, $A.\mathit{leftOrigin}$ and $B$ were consecutive, and $A$ was inserted between them. By (i), $B$'s left origin is not $A.\mathit{leftOrigin}$; thus, $B$ is not a descendant of $A.\mathit{leftOrigin}$ in the left-origin tree; and by \Cref{lem:fugue_left_origin}(b), $B$ is not a descendant of $A.\mathit{leftOrigin}$ in the FugueMax tree either. Therefore $A.\mathit{leftOrigin}$ did not have any right children in the FugueMax tree at the time when $A$ was inserted, and FugueMax's $\msf{insert}$ function made $A$ a right child of $A.\mathit{leftOrigin}$.

$C$ is also a right child of $A.\mathit{leftOrigin}$ in the FugueMax tree: it is a descendant by \Cref{lem:fugue_left_origin}(b); and if it were not a child, then its FugueMax parent would be a causally prior in-between element. Also, by assumption, $C$'s right origin is not $B$.

Thus $A$ and $C$ are right-side siblings in the FugueMax tree with different right origins. This situation is similar to elements $X$ and $Y$ in \Cref{fig:maximal-trees}. FugueMax's tree traversal then visits them in the reverse order of their right origins. Since $A \prec C$, this implies that $C$'s right origin appears before $B$ in the list order. But then $C$'s right origin is a causally prior in-between element, again contradicting our choice of $C$.

\textbf{Condition (3).}
Let $A$ and $B$ have the same left origin and the same right origin. Then FugueMax's $\msf{insert}$ function assigned them the same parent and side. FugueMax traverses these same-side siblings, with equal right origins, in order by ID.
\end{IEEEproof}

\subsection{Uniqueness of Maximally Non-Interleaving Order}
\label{sec:maximal_unique}
We finish by showing that maximal non-interleaving uniquely determines the list order. Specifically, any maximally non-interleaving replicated list induces the same total order on elements as FugueMax. This reinforces our claim that FugueMax is ``maximally'' non-interleaving---any additional nontrivial constraints would be impossible to satisfy.

\begin{theorem}
\label{thm:uniqueness}
Let $L$ be a replicated list algorithm that is maximally non-interleaving. Then $L$ is semantically equivalent to FugueMax. That is, in any execution, $L$ orders its list of elements according to FugueMax's global total order $\prec$ on elements.
\end{theorem}
\begin{IEEEproof}
We will show that conditions (1), (2), and (3) of \Cref{def:maximal} force $L$ to use a specific list order. Since FugueMax is maximally non-interleaving, this list order must be the same as FugueMax's.

First, condition (1) and \Cref{lem:rga_tree} imply that $L$'s list order is some depth-first pre-order traversal of the left-origin tree. Thus the only remaining degree of freedom for $L$'s list order is the order of siblings within the left-origin tree.

So, let $P$ be any element of $L$ or $\mathit{start}$, and let $S_P$ be the set of list elements whose left origin is $P$. Consider the forest of $S_P$ elements in which each element's parent is its right origin, except that an element has no parent if its right origin is not in $S_P$, or if its right origin is $\mathit{end}$.

For any parent-child edge $(A, B)$ in $S_P$, $A$ and $B$ have the same left origin ($P$). Thus $A$ and $B$ are \emph{not} an exception to condition (2). A mirrored version of the proof of \Cref{lem:rga_tree} shows that $L$'s list order, when restricted to any tree $T$ in $S_P$'s forest, must be a depth-first post-order traversal of $T$. (A depth-first post-order traversal satisfies the rule: traverse a node's children in some order, then visit the node.) Siblings within $T$ must be ordered by their IDs: they have the same left origins ($P$) and right origins (their parent), so condition (3) applies. Conditions (2) and (3) thus fully determine the order of elements in the same right-origin tree $T$.

It remains to specify the relative order of $S_P$ elements that are in different trees. We claim:
\begin{enumerate}[(a)]
    \item If $D$ and $E$ are tree roots in $S_P$ with $D \prec E$, then the entire tree rooted at $D$ (in $S_P$) appears before the entire tree rooted at $E$ in the list order.
    \item $L$ orders the tree roots in $S_P$ identically to how FugueMax sorts a node's right-side children: in the reverse order of their right origins, breaking ties by ID. (One can show that the tree roots in $S_P$ are precisely the right-side children of $P$ in FugueMax's tree.)
\end{enumerate}

To prove these claims, we first argue that all tree roots in $S_P$ were inserted concurrently. Indeed, in any state that already contains a tree root (more generally, an element with left origin $P$), the element $R$ immediately following $P$ must also have left origin $P$, because $L$'s list order is a depth-first pre-order traversal of the left-origin tree. A new element inserted into this state with left origin $P$ will have right origin $R \in S_P$, so it is not a tree root in $S_P$.

We prove the two claims above in turn:

\textbf{(a)} Let $A$ and $D$ be tree roots in $S_P$ such that $D \prec A$ in $L$, and let $B$ be a child of $A$. By the previous paragraph, $D$ is not causally later than $A$. Then the claim follows by a mirrored version of the last two paragraphs in \Cref{lem:rga_tree}'s proof (using condition (2) in place of forward non-interleaving).

\textbf{(b)} Let $D$ and $E$ be distinct tree roots in $S_P$. If $D$ and $E$ have the same right origins, then by condition (3), $L$ must order them by ID. Otherwise, let $R_D$ and $R_E$ be their right origins, and assume $R_E \prec R_D$. We need to show that $D \prec E$.

It is possible for a replica to be in a state $\sigma$ that contains $D$, $E$, and all causally prior elements, but no others. Since $E$ was inserted directly between $P$ and $R_E$, no elements causally prior to $E$ are between $P$ and $R_E$. Likewise, no elements causally prior to $D$ are between $P$ and $R_D$, hence none are between $P$ and $R_E \prec R_D$. Thus, $D$ and $E$ are the only elements in state $\sigma$ that can be between $P$ and $R_E$.

$D$ and $E$ are indeed between $P$ and $R_E$: because all tree roots in $S_P$ were inserted concurrently, $D$ and $E$ are the only elements in $\sigma$ with left origin $P$; thus they immediately follow $P$ in $L$'s depth-first pre-order traversal of the left-origin tree. So, state $\sigma$ contains either $\mathit{PDER_E}$ or $\mathit{PEDR_E}$ as a contiguous subsequence. Condition (2) forces $L$ to use $\mathit{PDER_E}$: $E$ is the last element in the list order with $R_E$ as its right origin, and there is no $C$ that could grant us an exception, so $E$ and $R_E$ must be consecutive. (By the same argument, \emph{YB} must be consecutive in \Cref{fig:maximal-noninterleaving}.) Hence $D \prec E$.
\end{IEEEproof}

Observe that the above proof gives an alternate characterization of FugueMax's list order:
\begin{enumerate}
    \item First, order elements by a depth-first pre-order traversal of the left-origin tree.
    \item Second, order siblings within that tree by a depth-first post-order traversal of their right-origin forest.
    \item Third, order roots within this forest that have different right origins by the reverse order of their right origins. (In Fugue, instead order them by ID.)
    \item Fourth, order roots within this forest that have the same right origin, and all other siblings in the forest, by ID.
\end{enumerate}

We chose to present FugueMax and Fugue as a double-sided tree for simplicity. But one could just as well use an algorithm based on the above characterization.

\section{Implementation and Evaluation}
\label{sec:impl}
We implemented two variations of Fugue, and one of FugueMax, in TypeScript. Each is written as a custom CRDT for the Collabs library \cite{collabs}; Collabs provides causal order delivery and other utilities.
The variations are:
\begin{itemize}
  \item \textbf{Fugue:} An optimized implementation of \Cref{alg:tree_fugue} in 1132 lines of code. It uses practical optimizations inspired by Yjs \cite{yjs_opts} and RGASplit \cite{rga_split}. In particular, it condenses sequentially-inserted tree nodes into a single ``waypoint'' object instead of one object per node, and it uses Protocol Buffers to efficiently encode update messages and saved documents. Collabs v0.6.1 uses this implementation for its list CRDTs.
  \item \textbf{Fugue Simple:} A direct implementation of \Cref{alg:tree_fugue} in 298 lines of code. It represents the state as a doubly-linked tree with one object per node, and it uses GZIP'd JSON encodings.
  \item \textbf{FugueMax Simple:} A direct implementation of \Cref{alg:tree_fugue} with FugueMax's modifications (see \Cref{sec:fugue_max_alg}) in 435 lines of code.
\end{itemize}

\subsection{Benchmarks}
We evaluated our implementations using a modified version of Jahns's crdt-benchmarks \cite{crdt_benchmarks}. Our Fugue and FugueMax implementations, benchmark code, and raw data are available on GitHub.\footnote{\url{https://github.com/mweidner037/fugue}}

Our goal with these benchmarks is to show that Fugue and FugueMax can be implemented with practical performance, comparable to existing CRDT implementations, in spite of the constraints imposed by non-interleaving. 

All experiments used Node.js v18.15.0 running on an Ubuntu 22.04.3
LTS desktop with a 4-core Intel i7 CPU @1.90GHz and 16GB RAM. For each metric, we performed 5 warmup trials followed by 10 measured trials; tables show mean $\pm$ standard deviation for the 10 measured trials.

We also compared to existing implementations in the crdt-benchmarks repository:
\begin{itemize}
  \item \textbf{Automerge-Wasm} (v0.5.0) is a CRDT with a JSON data model, implemented in Rust and compiled to WebAssembly.\footnote{\url{https://github.com/automerge/automerge}} Its list CRDT is based on RGA \cite{rga}.
  \item \textbf{Yjs} (v13.6.8) is a JavaScript CRDT library \cite{yjs}. Its list datatype is based on YATA \cite{yata}, and it is known for its good performance \cite{yjs_opts}.
  \item \textbf{Y-Wasm} (v0.16.10) is a Rust-to-WebAssembly variant of Yjs.\footnote{\url{https://github.com/y-crdt/y-crdt}}
\end{itemize}

Tables \ref{tab:saved_document} and \ref{tab:live} show results from a benchmark that replays a real-world text-editing trace \cite{kleppmann_text_trace}, in which every keystroke of the writing process for the {\LaTeX} source of a 17-page paper~\cite{Kleppmann2017json} was captured.
It consists of 182,315 single-character insert operations and 77,463 single-character delete operations, resulting in a final document size of 104,852 characters (not including tombstones).
Each implementation processed the full trace sequentially on a single replica.
Results for additional benchmarks, including microbenchmarks with concurrent operations, can be found in our GitHub repository.

\begin{table}
    \centering
    \caption{Saved document metrics. The plain text (without CRDT metadata or tombstones) is 105 kB in size.}
    \label{tab:saved_document}
    \begin{tabular}{lrrr} \toprule
         Implementation & Save size & Save time & Load time \\ \midrule
         Fugue & $168 \pm 0$ kB & $20 \pm 1$ ms & $13 \pm 2$ ms \\
         Fugue Simple & $1{,}021 \pm 0$ kB & $583 \pm 5$ ms & $334 \pm 3$ ms \\
         FugueMax Simple & $1{,}237 \pm 0$ kB & $788 \pm 11$ ms & $522 \pm 7$ ms \\
         Automerge-Wasm & $129 \pm 0$ kB & $180 \pm 0$ ms & $2{,}746 \pm 6$ ms \\
         Yjs & $160 \pm 0$ kB & $17 \pm 1$ ms & $63 \pm 6$ ms \\
         Y-Wasm & $160 \pm 0$ kB & $5 \pm 1$ ms & $15 \pm 0$ ms \\\bottomrule
    \end{tabular}
\end{table}

\Cref{tab:saved_document} considers the final saved document including CRDT metadata. In a typical collaborative app, this saved document would be saved (possibly on a server) at the end of each user session, and loaded at the start of the next session. Thus save size determines disk/network usage, while save/load time determines user-perceived save and startup latencies.

We see that Fugue is comparable to state-of-the-art Yjs on all three metrics, and the CRDT metadata is only 60\% of the literal text's size. Fugue Simple and FugueMax Simple are worse but still usable in practice. FugueMax Simple has a larger save size than Fugue Simple because it must additionally track right-side children's right origins.

\begin{table*}
    \centering
    \caption{Metrics for replaying a character-by-character text editing trace with 260k operations.}
    \label{tab:live}
    \begin{tabular}{lrrr} \toprule
         Implementation & Memory usage (MB) & Network bytes/op & Ops/sec (1,000s) \\ \midrule
         Fugue & $2.4 \pm 0.0$ & $46 \pm 0$ & $94 \pm 5$ \\
         Fugue Simple & $64.8 \pm 0.0$ & $151 \pm 0$ & $17 \pm 0$ \\
         FugueMax Simple & $71.9 \pm 0.0$ & $188 \pm 0$ & $16 \pm 0$  \\
         Automerge-Wasm & -- & $126 \pm 0$ & $52 \pm 0$ \\
         Yjs & $3.3 \pm 0.2$ & $29 \pm 0$ & $39 \pm 0$\\
         Y-Wasm & -- & $29 \pm 0$ & $7 \pm 0$ \\\bottomrule
    \end{tabular}
\end{table*}

\begin{table*}
    \centering
    \caption{Metrics for the real text trace repeated 100 times sequentially. The literal text has size 10,485 kB. We exclude implementations that use excessive time or memory.}
    \label{tab:text100}
    \begin{tabular}{lrrrr} \toprule
         Implementation & Save size (kB) & Save time (ms) & Load time (ms) & Memory usage (MB) \\ \midrule
         Fugue & $17{,}845 \pm 0$ & $1{,}405 \pm 35$ & $640 \pm 8$ & $223 \pm 0$ \\
         Yjs & $15{,}989 \pm 0$ & $479 \pm 35$ & $2{,}316 \pm 423$ & $294 \pm 17$ \\\bottomrule
    \end{tabular}
\end{table*}

\Cref{tab:live} shows performance metrics for live usage by a single user. Memory usage shows the increase in heap used\footnote{Measured with Node.js's \texttt{process.memoryUsage().heapUsed} after garbage collection. We exclude WebAssembly libraries because they do not use the JavaScript heap. (While resident set size measures WebAssembly memory usage, it had frequent outliers and gave implausible values for Automerge-Wasm.)} from the start to the end of the trace and thus approximates each list CRDT's in-memory size. Network bytes/op shows the average size of the per-op messages sent to remote collaborators. Ops/sec shows the average operation throughput; it reflects the time to process an op and encode a message for remote collaborators.
For example, Fugue achieves 94,000~ops/sec, an average of 11~{\textmu}s per operation.

We again see that Fugue is practical and comparable to Yjs. In particular, its memory usage is only a few MB---about 23 bytes per character, or 13 bytes per characters-including-tombstones. This refutes a common criticism of CRDTs for collaborative text editing: namely, that they have too much per-character memory overhead~\cite{Sun:2020}.
The memory overhead is worse for Fugue Simple (618 bytes/char) and FugueMax Simple (685 bytes/char),
but the total is still well within modern memory limits. For all Fugue variants, the network usage and operation throughput are far from being bottlenecks, given that a typical user types at $\approx 10$ chars/sec and a typical collaborative document has $<100$ simultaneous users.

\begin{figure}\centering
    \begin{subfigure}[t]{0.48\textwidth}
        \centering
        \includegraphics[width=\textwidth]{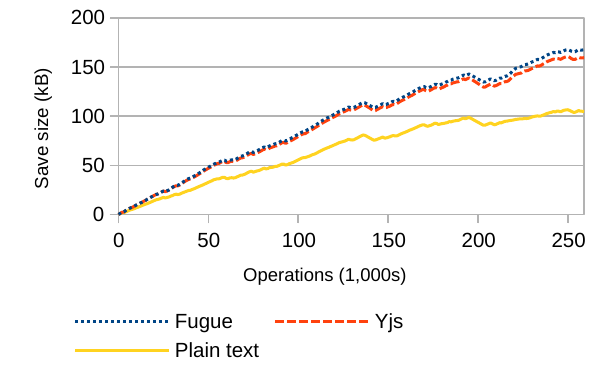}
        \caption{Saved document size.}
        \label{fig:doc_size}
    \end{subfigure}
    \begin{subfigure}[t]{0.48\textwidth}
        \centering
        \includegraphics[width=\textwidth]{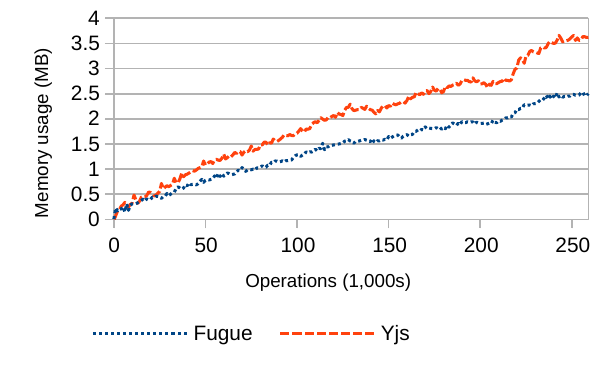}
        \caption{Memory usage.}
        \label{fig:mem_used}
    \end{subfigure}
    \caption{Metrics as a function of progress through the real text trace.}
\end{figure}

Figures \ref{fig:doc_size} and \ref{fig:mem_used} show how save size and memory usage vary throughout the text editing trace. The size of the plain text (without CRDT metadata or tombstones) is given for comparison. Observe that both metrics track the plain text's size at a modest multiple, and save size even decreases when text is deleted, despite tombstones.

Finally, \Cref{tab:text100} shows selected metrics for the same text-editing trace but repeated 100 times. The final document contains 10.5 million characters---several times longer than Tolstoy's \emph{War and Peace}. Nonetheless, Fugue's performance remains tolerable: 18MB save size, less than 2 seconds to save or load, and 223MB memory usage. Additionally, average network usage and throughput (not shown) remain within a $2\times$ factor of \Cref{tab:live}. 



\section{Conclusion}
Interleaving of concurrent insertions at the same position is an undesirable but largely ignored problem with many replicated list algorithms that are used for collaborative text editing. Indeed, most CRDT and OT algorithms that we surveyed exhibit interleaving anomalies. We also found that existing definitions of non-interleaving are impossible to satisfy.

In this paper, we proposed a new definition, maximal non-interleaving. We also introduced the Fugue and FugueMax list CRDTs and proved that FugueMax satisfies maximal non-interleaving, while Fugue is simpler and falls only slightly short of it. Our optimized implementation of Fugue has performance comparable to the state-of-the-art Yjs library.

In future work we plan to formally analyze Sync9 \cite{Sync9} and Gentle's modified version of the Yjs algorithm (YjsMod) \cite{gentle_code}. We conjecture that Sync9 is semantically equivalent to Fugue, while YjsMod is semantically equivalent to FugueMax. If so, then YjsMod is also maximally non-interleaving. These algorithms are also CRDTs; in future work it would be interesting to investigate whether maximally non-interleaving Operational Transformation algorithms exist.

\section*{Acknowledgments}
We thank Seph Gentle and Aryan Shah for insightful discussions and feedback on a draft of this paper. Matthew Weidner was supported by an NDSEG Fellowship sponsored by the US Office of Naval Research.
Martin Kleppmann's work was funded by the Volkswagen Foundation and crowdfunding supporters including SoftwareMill.

\bibliographystyle{IEEEtran}
\bibliography{references}

\begin{IEEEbiography}[{\includegraphics[width=1in,height=1.25in,clip,keepaspectratio]{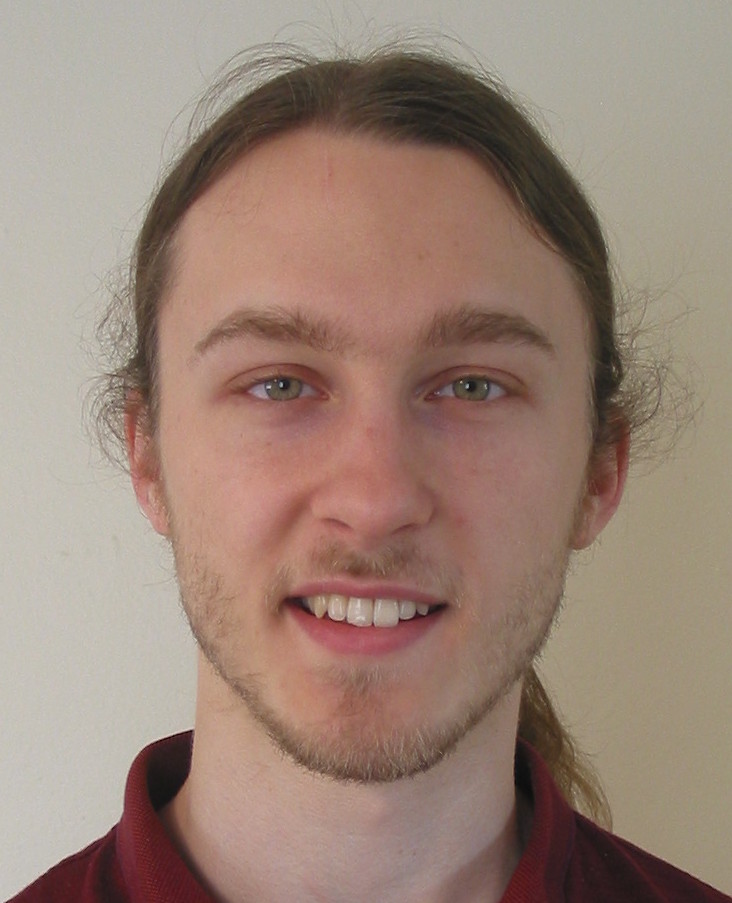}}]{Matthew Weidner}
is a PhD student at Carnegie Mellon University, advised by Heather Miller. He received the MPhil degree from the University of Cambridge, investigating decentralized cryptographic protocols. His research interests include distributed systems, data structures, and API design, with a focus on making CRDTs more flexible and easier to use.
\end{IEEEbiography}
\begin{IEEEbiography}[{\includegraphics[width=1in,height=1.25in,clip,keepaspectratio]{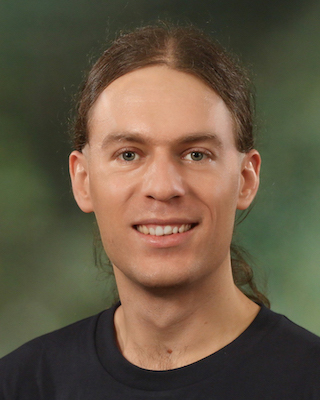}}]{Martin Kleppmann}
is an Associate Professor at the University of Cambridge. He focuses on decentralisation and distributed systems security, in particular local-first collaboration software, CRDTs, and formal verification. His book \emph{Designing Data-Intensive Applications} (O'Reilly Media) is the most popular text on data systems architecture, and it has been translated into eight languages. Previously, he was a software engineer and entrepreneur at Internet companies including Rapportive and LinkedIn.
\end{IEEEbiography}
\vfill

\ifincludeappendix
\appendices\newpage
\section{Problems with various algorithms}\label{sec:interleaving-examples}

\subsection{Examples of interleaving}
We give brief justifications of the interleaving claims ($\medblackcircle$) in \Cref{tab:survey}. Terminology and notation are as in the source cited for each algorithm.
For each algorithm we give a minimal example, in each case starting with an empty text document:
\begin{itemize}
    \item One user inserts $a$ followed by $b$, while concurrently another user inserts $x$.
    An algorithm exhibits forward interleaving (\Cref{def:forward}) if a possible merged outcome is $\mathit{axb}$.
    \item One user inserts $b$, and then prepends $a$ before $b$; concurrently another user inserts $x$.
    An algorithm exhibits backward interleaving if a possible merged outcome is $\mathit{axb}$.
\end{itemize}

\subsubsection{adOPTed and TTF forward interleaving}
adOPTed \cite{Ressel:1996} and TTF \cite{Oster:2006} use essentially the same transformation function for concurrent insertions.
We use the notation from TTF: an insertion operation is denoted $\mathit{ins}(p, c, s)$, where $p$ is the index at which to insert, $c$ is the inserted character, and $s$ is the ID of the site (i.e.,\ replica) on which the operation was generated.

To demonstrate forward interleaving, replica $A$ generates $\mathit{ins}(1,a,A)$ followed by $\mathit{ins}(2,b,A)$.
Concurrently, replica $B$ generates $\mathit{ins}(1,x,B)$.
Assume $A<B$ and consider the execution at $B$:
\begin{enumerate}
    \item $B$ executes $\mathit{ins}(1,x,B)$, so the document is $x$.
    \item When $B$ receives $\mathit{ins}(1,a,A)$, it computes \\
    $ T(\mathit{ins}(1,a,A), \mathit{ins}(1,x,B)) = \mathit{ins}(1,a,A)$ \\
    since $1=1 \wedge A<B$, so it inserts $a$ before $x$, yielding $\mathit{ax}$.
    \item When $B$ receives $\mathit{ins}(2,b,A)$, it computes \\
    $ T(\mathit{ins}(2,b,A), \mathit{ins}(1,x,B)) = \mathit{ins}(3,b,A) $ \\
    since $2>1$, so it inserts $b$ after $x$, yielding $\mathit{axb}$.
\end{enumerate}

\subsubsection{adOPTed/TTF backward interleaving (multi-replica)}

To demonstrate backward interleaving, we use three replicas, $A$, $B$, and $C$ with $A<B<C$.
First $C$ generates $\mathit{ins}(1,b,C)$, which is sent to $A$, and then $A$ generates $\mathit{ins}(1,a,A)$.
Concurrently, $B$ generates $\mathit{ins}(1,x,B)$.
Consider the execution at replica $B$:

\begin{enumerate}
    \item $B$ executes $\mathit{ins}(1,x,B)$, so the document is $x$.
    \item When $B$ receives $\mathit{ins}(1,b,C)$, it computes \\
    $ T(\mathit{ins}(1,b,C),\mathit{ins}(1,x,B)) = \mathit{ins}(2,b,C) $ \\
    since $1=1 \wedge C>B$, so it inserts $b$ after $x$, yielding $\mathit{xb}$.
    \item When $B$ receives $\mathit{ins}(1,a,A)$, it computes \\
    $ T(\mathit{ins}(1,a,A),\mathit{ins}(1,x,B)) = \mathit{ins}(1,a,A) $ \\
    since $1=1 \wedge A<B$, so it inserts $a$ before $x$, yielding $\mathit{axb}$.
\end{enumerate}

\subsubsection{Jupiter forward interleaving}

Jupiter \cite{Nichols:1995} is based on a single server that determines a total order of all operations, and it always transforms a client operation and a server operation with respect to each other.
When there are multiple clients and an operation from one client has been transformed and applied by the server, that transformed operation is considered a server operation from the point of view of the other clients (even though the operation actually originated on a client).

The Jupiter paper \cite{Nichols:1995} does not explicitly specify the transformation function for concurrent insertions, it only describes it verbally as: ``We arbitrarily chose to put server text first if both [i.e.\ server and client] try to insert at the same spot.''
We show that even this informal definition implies that the algorithm exhibits forward interleaving.

Starting with an empty document, assume that client $A$ generates $\mathit{ins}(1,a)$ followed by $\mathit{ins}(2,b)$, and concurrently client $B$ generates $\mathit{ins}(1,x)$.
Consider the execution at the server:

\begin{enumerate}
    \item Assume that client $A$'s $\mathit{ins}(1,a)$ is the first operation to reach the server, so the server simply applies it, resulting in the document $a$.

    \item Next, client $B$'s $\mathit{ins}(1,x)$ reaches the server.
    Since $\mathit{ins}(1,a)$ was concurrently applied by the server, we have two concurrent insertions at the same position.
    Per the rule for such insertions, the server's character, that is $a$, is placed first, and the client's $x$ second.
    This means $B$'s operation is transformed to $\mathit{ins}(2,x)$ and the server's document now reads $\mathit{ax}$.

    \item Finally, client $A$'s $\mathit{ins}(2,b)$ reaches the server.
    Concurrently, the server has now performed $\mathit{ins}(2,x)$, which is the transformed form of $B$'s operation.
    Again we have two concurrent insertions at the same position, in this case at index 2.
    Per the rule above we place the server's character $x$ first, and the client's character $b$ second.
    This means $A$'s operation is transformed to $\mathit{ins}(3,b)$ and the server's document now reads $\mathit{axb}$, exhibiting forward interleaving.
\end{enumerate}

If the rule is changed to place the client's insertion first and the server's insertion second in the case of concurrent insertions at the same position, the algorithm exhibits backward interleaving instead of forward interleaving.

\subsubsection{GOT forward interleaving}\label{sec:got-forward}

Strictly speaking, GOT is an OT control algorithm that is not specific to text; here we use GOT in conjunction with the transformation functions for text editing that are specified in the same paper \cite{Sun:1998:tochi}.
For insertions, those functions are an inclusion transformation $IT\_II$ and an exclusion transformation $ET\_II$.

To demonstrate forward interleaving, replica $A$ generates $\mathit{Insert}[a,1]$ followed by $\mathit{Insert}[b,2]$, while concurrently replica $B$ generates $\mathit{Insert}[x,1]$.
Let the total order on these operations be $\mathit{Insert}[a,1] < \mathit{Insert}[x,1] < \mathit{Insert}[b,1]$; this order is possible in the GOT timestamping scheme.
Consider the execution of operations at a replica that receives them in ascending order, so that we can skip the undo/do/redo process.
That replica will go through the following steps:
\begin{enumerate}
    \item Receive $\mathit{Insert}[a,1]$. We now have $HB = [\mathit{Insert}[a,1]]$, and the document is $a$.
    \item Receive $\mathit{Insert}[x,1]$.
    This operation is concurrent with all operations in $HB$, so it is transformed as follows:
    \begin{align*}
        EO_\mathit{new} &= LIT(\mathit{Insert}[x,1], HB[1,1]) \\
        &= IT\_II(\mathit{Insert}[x,1], \mathit{Insert}[a,1]) \\
        &= \mathit{Insert}[x,2]
    \end{align*}
    resulting in $HB=[\mathit{Insert}[a,1], \mathit{Insert}[x,2]]$ and the document $\mathit{ax}$.
    \item Receive $\mathit{Insert}[b,2]$.
    This operation is dependent on $HB[1]$ and concurrent with $HB[2]$, so it is transformed as follows:
    \begin{align*}
        EO_\mathit{new} &= LIT(\mathit{Insert}[b,2], HB[2,2]) \\
        &= IT\_II(\mathit{Insert}[b,2], \mathit{Insert}[x,2]) \\
        &= \mathit{Insert}[b,3]
    \end{align*}
    so $HB=[\mathit{Insert}[a,1], \mathit{Insert}[x,2], \mathit{Insert}[b,3]]$ and the document is $\mathit{axb}$.
\end{enumerate}

We discuss behavior of backward insertions in GOT in Appendix~\ref{sec:got-backward}.

\subsubsection{SOCT2, Logoot, and LSEQ}

Forward and backward interleaving in implementations of SOCT2~\cite{Suleiman:1997}, Logoot \cite{logoot}, and LSEQ \cite{lseq} are demonstrated in an open source repository~\cite{InterleavingTests}.

\subsubsection{WOOT forward interleaving}

The WOOT algorithm derives a partial order from the left and right origins of each inserted character, and then defines the document to be a unique linear extension of this partial order \cite{woot}.
The notation $\mathit{ins}(a \prec x \prec b)$ means that the character $x$ is inserted between $a$ and $b$.
$c_b$ marks the beginning and $c_e$ marks the end of the document.
When a replica receives an operation from another replica, a recursive function IntegrateIns determines where the insertion should be placed.

To demonstrate forward interleaving, replica $A$ generates $\mathit{ins}(c_b \prec a \prec c_e)$ to insert $a$, followed by $\mathit{ins}(a \prec b \prec c_e)$ to insert $b$.
Concurrently, $B$ generates $\mathit{ins}(c_b \prec x \prec c_e)$ to insert $x$.
Assume the ordering on the inserted characters' IDs is $a <_\mathit{id} x <_\mathit{id} b$.
Consider the execution at replica $B$:
\begin{enumerate}
    \item To apply $\mathit{ins}(c_b \prec x \prec c_e)$, calling $\mathrm{IntegrateIns}(x,c_b,c_e)$ results in the document $x$.
    \item To apply $\mathit{ins}(c_b \prec a \prec c_e)$, calling $\mathrm{IntegrateIns}(a,c_b,c_e)$ computes $S'=x$ and $L=c_b xc_e$.
    The loop stops at $i=1$ because $x >_\mathit{id} a$, so we recursively call the function $\mathrm{IntegrateIns}(a,c_b,x)$, resulting in the document $ax$.
    \item To apply $\mathit{ins}(a \prec b \prec c_e)$, calling $\mathrm{IntegrateIns}(b,a,c_e)$ computes $S'=x$ and $L=c_b xc_e$.
    The loop stops at $i=2$ because $x <_\mathit{id} b$, so we recursively call the function $\mathrm{IntegrateIns}(b,x,c_e)$, resulting in the document $\mathit{axb}$.
\end{enumerate}

\subsubsection{Treedoc forward and backward interleaving}

To demonstrate forward interleaving in Treedoc \cite{treedoc}, replica $A$ generates $\mathit{insert}(p_a, a)$ followed by $\mathit{insert}(p_b, b)$, while concurrently replica $B$ generates $\mathit{insert}(p_x, x)$.
These operations are assigned position identifiers $p_a=[(1:d_a)]$, $p_b=[1(1:d_b)]$, and $p_x=[(1:d_x)]$, where $d_a$, $d_b$, and $d_x$ are disambiguators.
Assume $d_a<d_x$; then the order on position identifiers is $p_a < p_x < p_b$, resulting in the document \emph{axb}.

To demonstrate backward interleaving in Treedoc, replica $A$ generates $\mathit{insert}(p_b, b)$ and then prepends $\mathit{insert}(p_a, a)$, while concurrently replica $B$ generates $\mathit{insert}(p_x, x)$.
These operations are assigned position identifiers $p_b=[(0:d_b)]$, $p_a=[0(0:d_a)]$, and $p_x=[(0:d_x)]$, where $d_b$, $d_a$, and $d_x$ are disambiguators.
Assume $d_x<d_b$; then the order on position identifiers is $p_a < p_x < p_b$, resulting in the document \emph{axb}.

\subsubsection{RGA backward interleaving}

To demonstrate backward interleaving in RGA~\cite{rga}, replica $A$ generates $\mathit{Insert}(1,b)$ followed by prepending $\mathit{Insert}(1,a)$, while concurrently replica $B$ generates $\mathit{Insert}(1,x)$.
The insertion of $b$ is assigned an s4vector of $\vec{v}_b=\langle 0,A,1,0 \rangle$, the insertion of $a$ an s4vector of $\vec{v}_a=\langle 0,A,2,0 \rangle$, and the insertion of $x$ an s4vector of $\vec{v}_x=\langle 0,B,1,0 \rangle$.
Assuming $A<B$, the order on these s4vectors is $\vec{v}_b \prec \vec{v}_x \prec \vec{v}_a$.
The left cobject (i.e., left origin) of all three operations is nil, and therefore the three elements are placed in their list in descending s4vector order, resulting in the document \emph{axb}.

\subsubsection{Yjs backward interleaving (multi-replica)}
\label{sec:yjs_interleave_code}
Yjs \cite{yjs} exhibits interleaving only in a limited case: when insertions occur in backward order across multiple replica IDs.
The following code demonstrates such interleaving in Yjs version 13.6.8. Our code repository contains a runnable copy of this code.

{\footnotesize
\begin{verbatim}
// Set up three replicas
const Y = require('yjs')
let doc1 = new Y.Doc(); doc1.clientID = 1
let doc2 = new Y.Doc(); doc2.clientID = 2
let doc3 = new Y.Doc(); doc3.clientID = 3

// Replica 3 inserts 'b'
doc3.getArray().insert(0, ['b'])

// Replica 1 inserts 'a' before 'b'
Y.applyUpdateV2(doc1,
  Y.encodeStateAsUpdateV2(doc3,
    Y.encodeStateVector(doc1)))
doc1.getArray().insert(0, ['a'])

// Replica 2 concurrently inserts 'x'
doc2.getArray().insert(0, ['x'])

// Prints the merged document: "axb"
Y.applyUpdateV2(doc1,
  Y.encodeStateAsUpdateV2(doc2,
    Y.encodeStateVector(doc1)))
console.log(doc1.getArray().toArray().join(''))
\end{verbatim}
}

\subsection{GOT character reordering}
\label{sec:got-backward}

We found that in the case of concurrent backward insertions, GOT exhibits an anomaly that is worse than interleaving: it reorders characters so that they appear in a different order from what the user typed, violating the replicated list specification.
In the following example, replica $A$ inserts $ab$ in backward order, while $B$ concurrently inserts $x$, and the resulting document reads $xba$~--- the $a$ and $b$ are reordered!
Like in Appendix~\ref{sec:got-forward}, we refer to the combination of the GOT control algorithm with the transformation functions specified in the same paper \cite{Sun:1998:tochi}.

Assume replica $A$ generates $\mathit{Insert}[b,1]$ followed by $\mathit{Insert}[a,1]$, while concurrently replica $B$ generates $\mathit{Insert}[x,1]$.
Let the total order on these operations be $\mathit{Insert}[x,1] < \mathit{Insert}[b,1] < \mathit{Insert}[a,1]$, which is consistent with causality.
Consider the execution of operations at a replica that receives them in ascending order.
That replica will go through the following steps:
\begin{enumerate}
    \item Receive $\mathit{Insert}[x,1]$. We now have the history buffer $HB = [\mathit{Insert}[x,1]]$, and the document is $x$.
    \item Receive $\mathit{Insert}[b,1]$.
    This operation is concurrent with all operations in $HB$, so it is transformed as follows:
    \begin{align*}
        EO_\mathit{new} &= LIT(\mathit{Insert}[b,1], HB[1,1]) \\
        &= IT\_II(\mathit{Insert}[b,1], \mathit{Insert}[x,1]) \\
        &= \mathit{Insert}[b,2]
    \end{align*}
    resulting in $HB=[\mathit{Insert}[x,1], \mathit{Insert}[b,2]]$ and the document $\mathit{xb}$.
    \item Receive $\mathit{Insert}[a,1]$. Now $HB[1]$ is concurrent to the new operation, but $HB[2]$ causally precedes it.
    Therefore $EOL = [\mathit{Insert}[b,2]]$ and
    \begin{align*}
        EOL' &= [LET(\mathit{Insert}[b,2], HB[1,1]^{-1})] \\
        &= [ET\_II(\mathit{Insert}[b,2], \mathit{Insert}[x,1])] \\
        &= [\mathit{Insert}[b,1]] \\
        O'_\mathit{new} &= LET(\mathit{Insert}[a,1], EOL'^{-1}) \\
        &= ET(\mathit{Insert}[a,1], \mathit{Insert}[b,1]) \\
        &= \mathit{Insert}[a,1] \\
        EO_\mathit{new} &= LIT(O'_\mathit{new}, HB[1,2]) \\
        &= IT\_II(IT\_II(\mathit{Insert}[a,1], \mathit{Insert}[x,1]), \\
        & \hspace{43pt}\mathit{Insert}[b,2]) \\
        &= IT\_II(\mathit{Insert}[a,2], \mathit{Insert}[b,2]) \\
        &= \mathit{Insert}[a,3]
    \end{align*}
    When $EO_\mathit{new}$ is applied, we obtain the incorrect document state $xba$.
\end{enumerate}

The GOT paper \cite[Definition 9]{Sun:1998:tochi} specifies that the transformation functions must satisfy a reversibility requirement, $IT(ET(O_a, O_b), O_b) = O_a$. However, the insert/insert transformation functions in Section 9 of the same paper do not meet this requirement.
Let $O_a = \mathit{Insert}[a,1]$ and $O_b = \mathit{Insert}[b,1]$. Then
\begin{align*}
&IT(ET(O_a, O_b), O_b) \\
&\quad = IT\_II(ET\_II(\mathit{Insert}[a,1], \mathit{Insert}[b,1]), \mathit{Insert}[b,1]) \\
&\quad = IT\_II(\mathit{Insert}[a,1], \mathit{Insert}[b,1]) \\
&\quad = \mathit{Insert}[a,2] \neq O_a
\end{align*}

We believe that this issue is not a typo or a similarly easy fix, but rather a deeper conceptual problem with the GOT algorithm.

\begin{figure}
    \centering
    \begin{tikzpicture}[font=\footnotesize]
        \tikzstyle{effector}=[draw,rectangle]
        \tikzstyle{generator}=[matrix,draw,rectangle,inner sep=2pt]
        \tikzstyle{network}=[thick,blue,-{Stealth[length=3mm]}]
        \node (left1) at (0,6) [effector] {\emph{``a''}};
        \node (right1) at (2.5,6) [effector] {\emph{``a''}};
        \node (left2) at (0,4.5) [generator] {
            \node {insert(1, $e$)};\\
            \node {$\rightarrow$ \emph{``ae''}};\\
        };
        \node [left,text width=2cm,align=right] at (left2.west) {$\mathit{ins}_e=$\\$(e,t_e,t_a,\emptyset)$};
        \node (right2) at (2.5,4.5) [generator] {
            \node {insert(1, $c$)};\\
            \node {$\rightarrow$ \emph{``ac''}};\\
        };
        \node [right,text width=2cm] at (right2.east) {$\mathit{ins}_c=$\\$(c,t_c,t_a,\emptyset)$};
        \node (left3) at (0,3) [generator] {
            \node {insert(1, $d$)};\\
            \node {$\rightarrow$ \emph{``ade''}};\\
        };
        \node [left,text width=2cm,align=right] at (left3.west) {$\mathit{ins}_d=$\\$(d,t_d,t_a,\{t_e\})$};
        \node (right3) at (2.5,3) [effector] {\emph{``ace''}};
        \node (left4) at (0,1.5) [effector] {\emph{``acde''}};
        \node (right4) at (2.5,1.5) [generator] {
            \node {insert(1, $b$)};\\
            \node {$\rightarrow$ \emph{``abce''}};\\
        };
        \node [right,text width=2cm] at (right4.east) {$\mathit{ins}_b=$\\$(b,t_b,t_a,\{t_c,t_e\})$};
        \node (left5) at (0,0) [effector] {?};
        \node (right5) at (2.5,0) [effector] {?};
        \draw [->] (left1) -- (left2);
        \draw [->] (left2) -- (left3);
        \draw [->] (left3) -- (left4);
        \draw [->] (left4) -- (left5);
        \draw [->] (right1) -- (right2);
        \draw [->] (right2) -- (right3);
        \draw [->] (right3) -- (right4);
        \draw [->] (right4) -- (right5);
        \draw [network] (left2.east) to [out=0,in=180] (right3.west);
        \draw [network] (left3.east) to [out=0,in=180] (right5.west);
        \draw [network] (right2.west) to [out=180,in=0] (left4.east);
        \draw [network] (right4.west) to [out=180,in=0] (left5.east);
    \end{tikzpicture}
    \caption{Execution that leads to divergence in Kleppmann et al.~\cite{interleaving}'s non-interleaving variant of RGA.}
    \label{fig:adithya-bug}
\end{figure}
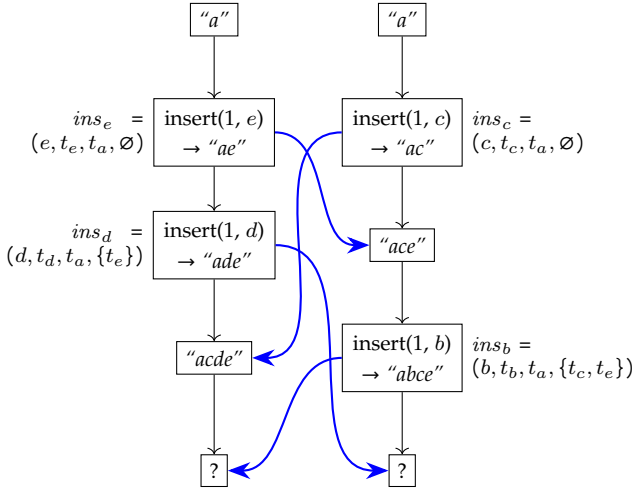

\subsection{Non-interleaving RGA does not converge}\label{sec:adithya-bug}

As explained in \Cref{sec:papoc-paper}, Kleppmann et al.~\cite{interleaving} previously attempted to design a non-interleaving text CRDT, but this algorithm does not converge.
We now give an example of this problem, which was found by Chandrassery~\cite{adithya}.

The algorithm is a variant of Attiya et al.~\cite{attiya}'s timestamped insertion tree (which is a reformulation of RGA~\cite{rga}).
Each insertion operation includes the ID of a \emph{reference element} (after which the new character should be inserted), and additionally this algorithm includes the set of existing characters with the same reference element.
When determining the order of characters with the same reference element, this additional metadata is taken into account.

The correctness of the algorithm depends on a strict total ordering relation $<$ that determines the order in which characters with the same reference element should appear in the document.
However, it is possible to construct executions in which three insertion operations $x$, $y$, and $z$ are ordered $x<y$, $y<z$, and $z<x$, violating the asymmetry property of the total order.
If $<$ is not a total order, the order of characters in the document is ambiguous, and so the algorithm cannot guarantee that replicas converge to the same state.

\Cref{fig:adithya-bug} shows an execution that triggers the problem.
Each insertion operation is a tuple $(a,t,r,e)$ where $a$ is the character being inserted, $t$ is the timestamp (unique ID) of the operation, $r$ is the timestamp of the reference character (immediate predecessor at the time the operation was generated), and $e$ is the set of siblings (operations with the same reference character at the time the operation was generated).

Assume $t_e < t_c < t_d < t_b$.
To determine the total order of operations, we first apply the rule that $\mathit{op}_1 < \mathit{op}_2$ if $\mathit{op}_1$'s timestamp appears in $\mathit{op}_2$'s siblings, which gives us:
\begin{align*}
    \mathit{ins}_e = (e,t_e,t_a,\emptyset) &< (d,t_d,t_a,\{t_e\}) = \mathit{ins}_d \\
    \mathit{ins}_c=(c,t_c,t_a,\emptyset) &< (b,t_b,t_a,\{t_c,t_e\}) = \mathit{ins}_b \\
    \mathit{ins}_e = (e,t_e,t_a,\emptyset) &< (b,t_b,t_a,\{t_c,t_e\}) = \mathit{ins}_b
\end{align*}

Next, we compare $\mathit{ins}_b$ and $\mathit{ins}_d$ using the rule for concurrent operations:
\begin{gather*}
    \min(\{t_b\} \cup \{t_c,t_e\} - \{t_e\}) = t_c \\
    \min(\{t_d\} \cup \{t_e\} - \{t_c,t_e\}) = t_d \\
    t_c < t_d, \text{ therefore: } \mathit{ins}_b < \mathit{ins}_d
\end{gather*}

Finally, we compare $\mathit{ins}_c$ and $\mathit{ins}_d$ using the rule for concurrent operations:
\begin{gather*}
    \min(\{t_c\} \cup \emptyset - \{t_e\}) = t_c \\
    \min(\{t_d\} \cup \{t_e\} - \emptyset) = t_e \\
    t_e < t_c, \text{ therefore: } \mathit{ins}_d < \mathit{ins}_c
\end{gather*}

We now have $\mathit{ins}_c < \mathit{ins}_b$, $\mathit{ins}_b < \mathit{ins}_d$, and $\mathit{ins}_d < \mathit{ins}_c$.
This violates the requirement that $<$ is a total order.
Therefore, the order of characters in the document is not uniquely determined, and we cannot guarantee that replicas will converge to the same state.

\fi 
\end{document}